\documentclass[aps,superscriptaddress]{revtex4-1}
\pdfoutput=1

\usepackage{amsmath}
\usepackage{graphicx}
\usepackage{bbm}

\begin{document}

\newcommand{\bra}[1] {\left\langle #1 \right|}
\newcommand{\ket}[1] {\left| #1 \right\rangle}

\title{Selective darkening of degenerate transitions for implementing quantum controlled-NOT gates}

\author{P. C. de Groot}
\email{pieter.degroot@lmu.de}
\affiliation{Kavli Institute of Nanoscience, Delft University of Technology, Post Office Box 5046, 2600 GA Delft, The Netherlands}
\affiliation{Max Planck Institute for Quantum Optics, Garching 85748, Munich, Germany}

\author{S. Ashhab}
\affiliation{Advanced Science Institute, RIKEN, Wako-shi, Saitama 351-0198, Japan}
\affiliation{Department of Physics, University of Michigan, Ann Arbor, Michigan 48109-1040, USA}

\author{A.~Lupa\c{s}cu}
\affiliation{Institute for Quantum Computing, Waterloo Institute for Nanotechnology, and Department of Physics and Astronomy, University of Waterloo, N2L 5G7 Waterloo, Canada}

\author{L. DiCarlo}
\affiliation{Kavli Institute of Nanoscience, Delft University of Technology, Post Office Box 5046, 2600 GA Delft, The Netherlands}

\author{Franco Nori}
\affiliation{Advanced Science Institute, RIKEN, Wako-shi, Saitama 351-0198, Japan}
\affiliation{Department of Physics, University of Michigan, Ann Arbor, Michigan 48109-1040, USA}

\author{C. J. P. M. Harmans}
\affiliation{Kavli Institute of Nanoscience, Delft University of Technology, Post Office Box 5046, 2600 GA Delft, The Netherlands}

\author{J. E. Mooij}
\affiliation{Kavli Institute of Nanoscience, Delft University of Technology, Post Office Box 5046, 2600 GA Delft, The Netherlands}

\date{\today}

\begin{abstract}
  We present a theoretical analysis of the selective darkening method for implementing quantum controlled-NOT (CNOT) gates. This method, which we recently proposed and demonstrated, consists of driving two transversely-coupled quantum bits (qubits) with a driving field that is resonant with one of the two qubits.
  For specific relative amplitudes and phases of the driving field felt by the two qubits, one of the two transitions in the degenerate pair is darkened, or in other words, becomes forbidden by effective selection rules. At these driving conditions, the evolution of the two-qubit state realizes a CNOT gate.
  The gate speed is found to be limited only by the coupling energy \( J \), which is the fundamental speed limit for any entangling gate. Numerical simulations show that at gate speeds corresponding to 0.48$J$ and 0.07$J$, the gate fidelity is 99\% and 99.99\%, respectively, and increases further for lower gate speeds. In addition, the effect of higher-lying energy levels and weak anharmonicity is studied, as well as the scalability of the method to systems of multiple qubits. We conclude that in all these respects this method is competitive with existing schemes for creating entanglement, with the added advantages of being applicable for qubits operating at fixed frequencies (either by design or for exploitation of coherence sweet-spots) and having the simplicity of microwave-only operation.
\end{abstract}
%


\maketitle

\section{Introduction}
  An important step towards the realisation of a quantum computer is to implement a set of universal gates from which all other operations can be composed. The quantum controlled-NOT (CNOT) gate is a prominent example of a two-qubit universal gate, requiring only the addition of single-qubit gates to form a complete universal set.
  The physical implementation of a gate depends on the type of qubit that is used, but even for a specific qubit type a certain gate can often be implemented using a variety of methods. The ideal method delivers a gate that is fast compared to qubit decoherence times, has a high fidelity, is simple to implement and does not introduce constraints that compromise the coherence time of the qubits or the implementation of other gates.

  One classification of two-qubit gates is based on the required nature of the interaction between the qubits \cite{geller:10}. When the interaction term contains diagonal elements in the single-qubit energy eigenbasis, often referred to as longitudinal or \( zz \)-coupling, the transition frequency of one qubit depends on the state of the other qubit. Although this spectroscopic shift enables simple resonant driving for all operations \cite{linden:98, plantenberg:07}, the shift also leads to continuously evolving phases that must be compensated by refocusing schemes \cite{jones:99}.
  In contrast, for transverse coupling the energy splitting of one qubit does not depend on the state of the other qubit. In this respect, the system can be described as a set of effectively uncoupled qubits \cite{rigetti:05, paraoanu:06}. The desired entangling evolution can then be induced by dynamic manipulation of the system. Clearly, the degeneracy due to the lack of a spectroscopic splitting forbids any method based on frequency-selectivity.
  Focusing on superconducting qubits \cite{makhlin:01, you:05, wendin:07, clarke:08, schoelkopf:08, shevchenko:10, ladd:10, you:11, buluta:11},
  previous methods used either additional coupling elements \cite{averin:03, izmalkov:04, plourde:04, maassen:05, wallquist:05, bertet:06, hime:06, hutter:06, niskanen:07, sillanpaa:07, yamamoto:08, helmer:09, harrabi:09, wang:09},
  extra qubit states outside the computational basis \cite{strauch:03, sillanpaa:07, neeley:10, dicarlo:10}
  or shifting levels in and out of resonance by DC \cite{yamamoto:03, berkley:03, mcdermott:05}
  or strong AC fields \cite{rigetti:05, ashhab:06, ashhab:07, majer:07}.

  Another important consideration is that a method can require certain bias conditions of the qubits, which might not always be compatible with optimal coherence time conditions,
  or the implementations of other gates. This is especially important for qubits with a so-called sweet-spot, an optimal bias point at which the qubit is rendered insensitive to specific noise channels, and coherence times can improve by orders of magnitude. For some qubit types and coupling schemes these sweet-spots are intrinsically connected to having a transverse coupling. Previous work \cite{rigetti:05, ashhab:06, paraoanu:06, bertet:06, groot:10a, rigetti:10, chow:11} has specifically focused on two-qubit quantum gates that can be fully implemented at sweet-spots.

  In this work we theoretically analyse a CNOT gate based on the selective darkening method as proposed and demonstrated in reference \cite{groot:10a}. This method is for transversely-coupled qubits and requires driving two qubits simultaneously with a single frequency, with specific relative amplitudes and phases. The method was developed in the context of superconducting qubits, but can also be used for any other coupled-qubit system with pure transverse coupling \cite{wei:11}. The basic principle of the gate is explained in section~\ref{sec:selda}, for both small and large coupling energies. In section~\ref{sec:othergates} we briefly discuss the relation of selective darkening to other comparable methods. The time-domain evolution of the gate is investigated in section~\ref{sec:timedom}, including the possible gate errors caused by ac-Stark shifts due to off-resonant driving of other transitions in the system. A numerical study of the gate evolution, gate errors and gate speed is presented in section~\ref{sec:num2level}. The influence of possible higher-lying levels is analysed for weakly anharmonic qubits in section~\ref{sec:weakanharmonic}. Lastly, the scalability of the method to systems of multiple qubits is studied in section~\ref{sec:scalability}, and we finish with the conclusions and discussion of the results.

\section{Selective darkening}\label{sec:selda}

    \subsection{Basic principle}\label{sec:seldaA}

    Selective darkening provides an entangling operation between two coupled qubits. To operate the gate, one drives both qubits with a common frequency, but individually tuned amplitudes and phases. In this section we first give an intuitive explanation of the method, and for simplicity assume a coupling energy between the qubits that is much smaller than the other energies in the system.

    The selective darkening method is suitable for any coupled-qubit system where the effective coupling term is purely transverse, i.e. where the matrix describing the interaction, written in the energy eigenbasis of the uncoupled system, does not contain any diagonal elements. Here we consider a system of two transversely coupled qubits that is described by the Hamiltonian
  \begin{equation}
    \hat{H}_0 = - \frac{1}{2} \left( \Delta_1 \hat{\sigma}_z^{(1)} + \Delta_2 \hat{\sigma}_z^{(2)} \right) + J \hat{\sigma}_x^{(1)} \hat{\sigma}_x^{(2)},
    \label{eq:Hamiltonian}
  \end{equation}
  where \( \Delta_i \) is the single-qubit energy splitting of qubit \( i \), \( J \) is the qubit-qubit coupling energy and \( \hat{\sigma}_{x,y,z}^{(i)} \) are the Pauli spin matrices (\( \hat{\sigma}_{x,y,z}^{(1)} = \hat{\sigma}_{x,y,z} \otimes \mathbbm{1} \), \( \hat{\sigma}_{x,y,z}^{(2)} = \mathbbm{1} \otimes \hat{\sigma}_{x,y,z} \)).
  Note however that other (effective) Hamiltonian with pure transverse coupling, i.e. with any combination of \( \hat{\sigma}_{x,y}^{(1)} \hat{\sigma}_{x,y}^{(2)} \) coupling terms, would not change the principle of the method. The same remark holds for systems where the coupling is not direct between the two qubits, but mediated by an extra coupling element such as a harmonic oscillator \cite{blais:07} or a SQUID \cite{hime:06} (Superconducting QUantum Interference Device). In figures~1(a) and 1(b) two relevant example systems are depicted.

    \begin{figure}[htbp]
      \begin{center}
        \includegraphics{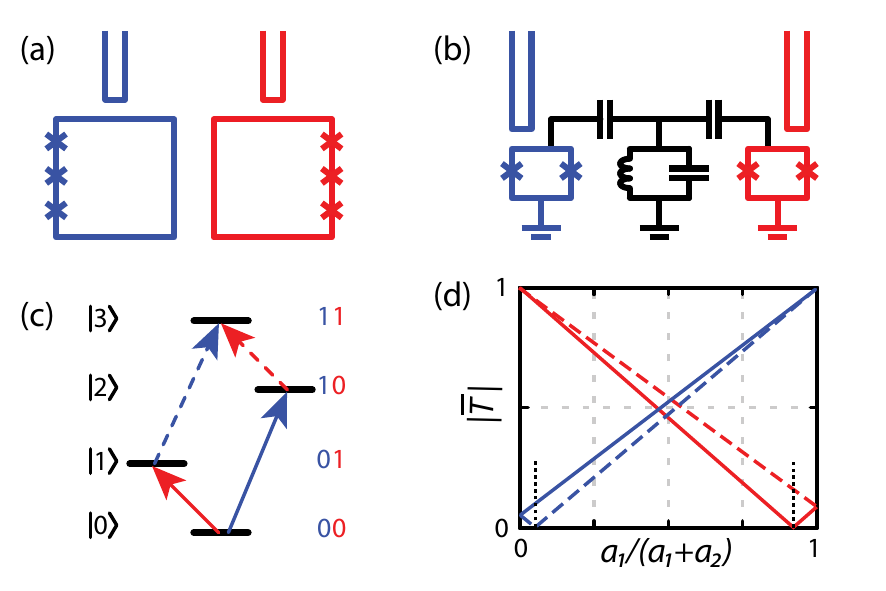}
      \end{center}
      \caption{Selective darkening
      (a) Schematic diagram of two inductively coupled flux qubits and two antennas for individual driving.
      (b) Schematic diagram of two transmon or charge qubits, coupled via a harmonic oscillator, and with two antennas for individual driving.
      (c) Energy level diagram of the coupled-qubit system. Arrows of the same color indicate
      transitions of the same qubit and are degenerate in frequency.
      (d) The normalized transition strengths of the four transitions in (c) as a function of the ratio of driving amplitudes \( a_1/(a_1+a_2) \) for \( \varphi_2-\varphi_1=0 \). For \( \varphi_2-\varphi_1=\pi \) the dashed and solid lines are interchanged. The black dotted lines indicate the condition for which the corresponding transition is darkened.
}
      \label{fig:fig1}
    \end{figure}

  As a first step we treat the coupling strength $J$ as a
perturbation, taking \( J \ll |\Delta_1-\Delta_2| \). We also assume that
$|\Delta_1-\Delta_2|\ll\Delta_1,\Delta_2$, so that we can ignore
terms containing the ratio $J/\Delta_j$. With no loss of
generality, we also take $\Delta_1>\Delta_2$. To first order
in perturbation theory, the four energy eigenstates are given by:
\begin{eqnarray}
\ket{0} & = & \ket{00} \nonumber
\\
\ket{1} & = & \ket{01} - \frac{J}{\Delta_1-\Delta_2} \ket{10} \nonumber
\\
\ket{2} & = & \ket{10} + \frac{J}{\Delta_1-\Delta_2} \ket{01} \nonumber
\\
\ket{3} & = & \ket{11}.
\end{eqnarray}
The energies are unaffected by the perturbation to first order. In other words
$E_2-E_0=E_3-E_1=\Delta_1$ and $E_1-E_0=E_3-E_2=\Delta_2$.
(In the next section it is shown that the relations $E_2-E_0=E_3-E_1$ and
$E_1-E_0=E_3-E_2$ hold to \emph{any} order).
  The energy levels are shown schematically in figure~1(c). The arrows indicate the transitions of interest; the blue and red arrows describe the transitions of qubit 1 and 2, respectively. As mentioned above, each pair of transitions is degenerate, which is typical for transverse coupling
  \footnote{For simplicity, we label the states as if the qubits were uncoupled, although the single-qubit states are mixed by the coupling. This state mixing is central to the selective darkening method. Note that in the case of fixed coupling, this state mixing needs to be taken into account when defining a suitable readout scheme.}.
Let us now add a driving term of frequency \( \omega \), resonant with one of the degenerate pairs of transitions, that couples with amplitude \( a_1 \) and phase \( \varphi_1 \) to qubit~1, and \( a_2 \) and \( \varphi_2 \) to qubit~2:
\begin{eqnarray}
\hat{H}_{\rm drive} & = & a_1 \cos (\omega t + \varphi_1)
\hat{\sigma}_x^{(1)} + a_2 \cos (\omega t + \varphi_2) \hat{\sigma}_x^{(2)} \nonumber
\\
& = & \tilde{H}_{\rm drive}^{+} e^{i\omega t} + \tilde{H}_{\rm drive}^{-} e^{-i\omega t},
\label{eq:Hdrive}
\end{eqnarray}
where
\begin{equation}
\tilde{H}_{\rm drive}^{\pm} =
\frac{a_1}{2} e^{\pm i\varphi_1} \hat{\sigma}_x^{(1)} +
\frac{a_2}{2} e^{\pm i\varphi_2} \hat{\sigma}_x^{(2)}.
\end{equation}
The amplitudes \( a_1 \) and \( a_2 \) are real and positive, and the phases \( \varphi_1 \) and \( \varphi_2 \) are real. If we choose $\omega=\Delta_2/\hbar$ (in our first-order approximation),
we match the transition frequency for flipping the state of the
second qubit, i.e. we should drive the transitions $\ket{0}
\leftrightarrow \ket{1}$ and $\ket{2} \leftrightarrow \ket{3}$ [corresponding to red arrows in figure~1(c)].
We now evaluate the transition strengths \( T_{k \leftrightarrow l} = \bra{l} \tilde{H}_{\rm drive} \ket{k} \) of both transitions, i.e. the matrix elements that govern the Rabi frequencies of the oscillations (\( \omega_R = 2|T|/\hbar \)). The term \( \tilde{H}_{\rm drive} \) represents the (time-independent) co-rotating field, meaning that we take the rotating wave approximation, giving \( \tilde{H}_{\rm drive} = \tilde{H}_{\rm drive}^{+} \) for \( k > l \) and \( \tilde{H}_{\rm drive} = \tilde{H}_{\rm drive}^{-} \) for \( k < l \) (The derivation of the co- and counter-rotating terms is done in appendix~\ref{apx:rotframe}). The resulting transition strengths are:
\begin{subequations}
\begin{eqnarray}
\bra{1} \tilde{H}_{\rm drive} \ket{0} & = &
- \frac{a_1}{2} \frac{J}{\Delta_1-\Delta_2} e^{-i\varphi_1}
+ \frac{a_2}{2} e^{-i\varphi_2}
\\
\bra{3} \tilde{H}_{\rm drive} \ket{2} & = &
+ \frac{a_1}{2} \frac{J}{\Delta_1-\Delta_2} e^{-i\varphi_1}
+ \frac{a_2}{2} e^{-i\varphi_2}.
\end{eqnarray}
\end{subequations}
Crucially, the two transition strengths are not equal. Figure~1(d) shows the normalized \( |\overline{T}| = |T|/(a_1+a_2) \) as a function of \( a_1/(a_1+a_2) \), for a fixed phase difference \( \varphi_2-\varphi_1 = 0 \). The blue lines correspond to the two transitions of qubit 2.
The difference in transition strength leads to a different evolution for the two corresponding transitions. This can be conveniently visualized using a Bloch sphere representation, as shown in figure~2(a). The black and grey arrows both represent the state of the qubit that is resonantly driven (in our example qubit 2), where black (grey) indicates that the other qubit is in state \( \ket{0} \) (\( \ket{1} \)).
The transition strength is a complex number and both the amplitude and phase are relevant for the evolution of the state. Difference in the \emph{amplitudes} of the two transition strengths leads to different Rabi frequencies of the oscillations (\( \omega_{R,1} \neq \omega_{R,2} \)). Difference in the \emph{phases} of the transition strengths leads to different rotation axes (\( \varphi_{R,1} \neq \varphi_{R,2} \)).

We make some simple observations. A single-qubit gate is a gate that changes the state of one qubit with the change being independent of the state of the other qubit(s). In the Bloch sphere picture this means that when a rotation is induced on one qubit, the rotation speed and the orientation of the rotation axis do not depend on the other qubit being in state \( \ket{0} \) or \( \ket{1} \). This situation is indicated in figure~2(d). This only occurs when the driving field couples exclusively to the qubit that is resonant with the driving field; in our case this corresponds to \( a_1=0 \) (or \( a_1/(a_1+a_2)=0 \), see figure~1(c)). \emph{All other settings of the driving field lead to entangling evolution of the system, and can be used to create entangling gates.} Note that if one intends to perform a single-qubit gate, and the drive is not applied exclusively to that qubit, the two-qubit evolution that is aimed for in this work should be taken into account as a possible source of errors, unless the system has a tunable coupling that can be reduced to zero~\cite{goszkowski:11, averin:03, niskanen:07}.

    \begin{figure}[htbp]
      \begin{center}
        \includegraphics{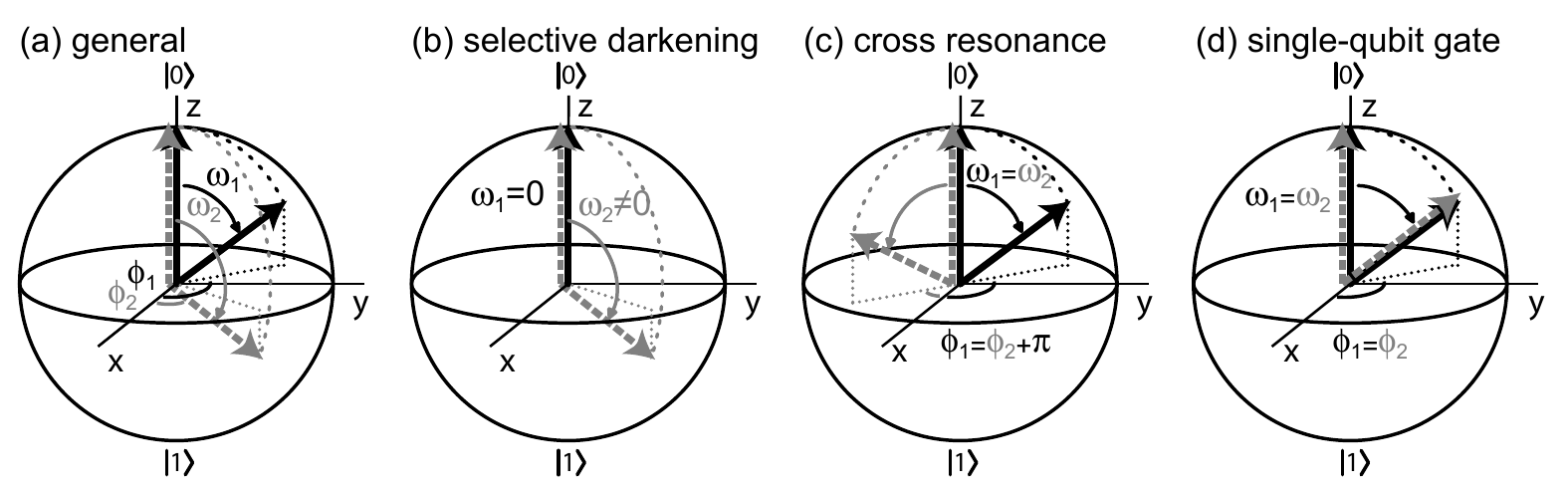}
      \end{center}
      \caption{
      Bloch sphere representation of the evolution of a resonantly driven degenerate transition for different driving conditions.
      The black and grey arrows show the evolution of the state of the target qubit, when the control qubit is in state \( \ket{0} \) and state \( \ket{1} \), respectively.
      (a) General case. The two state vectors rotate with two different Rabi frequencies, and around two different axes.
      (b) Selective darkening. The rotation of one of the state vectors is fully suppressed. The other state-vector has a non-zero Rabi frequency.
      (c) Cross resonance. The state vectors rotate around the same axis, with equal Rabi frequency, but in opposite directions.
      (d) Single-qubit gate. The state vectors rotate around the same axis, with equal Rabi frequency and in the same direction.}
      \label{fig:bloch}
    \end{figure}

Even though many different settings of the driving field can be used to create entangling gates, two special cases stand out. The first is where the driving field resonant with one qubit is exclusively applied to the non-resonant qubit (\( a_2=0  \), \( a_1/(a_1+a_2)=1 \)). This driving scenario will lead to equal rotation speed of both state vectors, but opposite direction. This case will be discussed further in section~\ref{sec:othergates}.

  The second special case is the main focus of this paper, where one of the transition strengths equals zero [depicted in figure~1(d) by the black dotted lines]. Here one transition is fully suppressed, i.e. darkened, while the other transition has a finite transition strength.
Specifically, if we choose:
\begin{equation}
\frac{a_2}{a_1} = \frac{J}{\Delta_1-\Delta_2},
\label{eq:AmplitudeRatio}
\end{equation}
and $\varphi_2-\varphi_1=0$, we find that the Rabi oscillation frequency for
the transition $\ket{0} \leftrightarrow \ket{1}$ is zero, so that
we only drive the transition $\ket{2} \leftrightarrow \ket{3}$.
This condition provides exactly what is required for a CNOT gate: for the appropriate driving duration the state of the second qubit is flipped
if the first qubit is in state $\ket{1}$, while it remains unaffected if the first qubit is in state \( \ket{0} \). The 0-CNOT gate,
where the second qubit is flipped only if the first qubit is in state \( \ket{0} \),
is achieved by taking equation~(\ref{eq:AmplitudeRatio}) and $\varphi_1-\varphi_2=\pi$.

The speed of the gate is determined by the transition strength of
the non-darkened transition, which
is now given by:
\begin{equation}
  \bra{3} \tilde{H}_{\rm drive} \ket{2} = a_2 = a_1 \frac{J}{\Delta_1-\Delta_2}.
\label{eq:transstr}
\end{equation}
One must take care that both $a_1$ and $a_2$ are small enough to not excite other transitions by off-resonant excitation. As a simple estimation for the maximum gate speed we take the driving amplitudes such that for all the off-resonant transitions the transition strength is smaller than the detuning of the transition frequency with the driving field: \( \bra{l} \tilde{H}_{\rm drive} \ket{k} < |\hbar\omega - \hbar\omega_{l \leftrightarrow k}| \). In this case the transitions $\ket{0} \leftrightarrow \ket{2}$ and $\ket{1} \leftrightarrow \ket{3}$ provide the tightest restrictions (\( \bra{2} \tilde{H}_{\rm drive} \ket{0} \approx \bra{3} \tilde{H}_{\rm drive} \ket{1} \approx a_1/2 \)), giving \( a_1/2 < (\Delta_1 - \Delta_2) \). The resulting estimate for the maximum Rabi frequency of the CNOT transition is
\begin{equation}
  \omega_{R,{\rm max}} \sim 4J / \hbar.
\end{equation}
This means that the maximum
gate speed is determined by $J$, which is the
fundamental upper limit on performing a two-qubit gate;
a two-qubit gate cannot be faster than the coupling strength in
the system \cite{ashhab:12}.
How close the gate can get to this estimate for the
maximum speed (for a given fidelity) is studied numerically in section~\ref{sec:num2level}.

  \subsection{Strong coupling}\label{sec:seldaB}

In the simplified calculations of the previous subsection, the energy eigenstates
are easily identifiable as being almost equal to the states of the
computational basis. However, the Hamiltonian [equation~(\ref{eq:Hamiltonian})] is simple enough that
it can be diagonalized without approximations:
\begin{eqnarray}
\ket{0} & = & \cos(\theta_1/2) \ket{00} - \sin(\theta_1/2)
\ket{11} \nonumber
\\
\ket{1} & = & \cos(\theta_2/2) \ket{01} - \sin(\theta_2/2)
\ket{10} \nonumber
\\
\ket{2} & = & \cos(\theta_2/2) \ket{10} + \sin(\theta_2/2)
\ket{01} \nonumber
\\
\ket{3} & = & \cos(\theta_1/2) \ket{11} + \sin(\theta_1/2)
\ket{00}, \label{eq:eigv2}
\end{eqnarray}
where
\begin{equation}
  \tan \theta_1 = \frac{2J}{\Delta_1+\Delta_2}, \hspace{12pt} \tan \theta_2 = \frac{2J}{\Delta_1-\Delta_2}.
\end{equation}
For compactness of the following equations, we also define
\begin{equation}
  \theta_+ = \frac{\theta_1 + \theta_2}{2}, \hspace{12pt} \theta_- = \frac{\theta_2 - \theta_1}{2}.
\end{equation}
We now find that
\begin{subequations} \label{eq:transstr_strongJ}
\begin{eqnarray}
\bra{1} \tilde{H}_{\rm drive} \ket{0} & = &
- \frac{a_1}{2} e^{-i\varphi_1} \sin\theta_+
+ \frac{a_2}{2} e^{-i\varphi_2} \cos\theta_-
\\
\bra{3} \tilde{H}_{\rm drive} \ket{2} & = &
+ \frac{a_1}{2} e^{-i\varphi_1} \sin\theta_+
+ \frac{a_2}{2} e^{-i\varphi_2} \cos\theta_-
\end{eqnarray}
and
\begin{eqnarray}
\bra{2} \tilde{H}_{\rm drive} \ket{0} & = &
\frac{a_1}{2} e^{-i\varphi_1} \cos\theta_+ +
\frac{a_2}{2} e^{-i\varphi_2} \sin\theta_-
\\
\bra{3} \tilde{H}_{\rm drive} \ket{1} & = &
  \frac{a_1}{2} e^{-i\varphi_1} \cos\theta_+
- \frac{a_2}{2} e^{-i\varphi_2} \sin\theta_-.
\end{eqnarray}
\end{subequations}
The condition for selective darkening of qubit 2 transitions is then given by
\begin{equation}
\frac{a_2}{a_1} =
\frac{\sin\theta_+}{\cos\theta_-},
\label{eq:sd_condition_strongdr}
\end{equation}
with $\varphi_2-\varphi_1=0$ for the 1-controlled CNOT gate (1-CNOT), and $\varphi_2-\varphi_1=\pi$
for the 0-CNOT gate. This condition generalizes equation~(\ref{eq:AmplitudeRatio}), which was derived for the weak-coupling limit. The condition for selective darkening of qubit 1 transitions is
\begin{equation}
\frac{a_2}{a_1} =
\frac{\cos\theta_+}{\sin\theta_-},
\end{equation}
with $\varphi_2-\varphi_1=\pi$ for the 1-CNOT gate, and $\varphi_2-\varphi_1=0$
for the 0-CNOT gate.
In this case (i.e., with arbitrary values of $\Delta_1$, $\Delta_2$
and $J$), the transition frequencies are given by:
\begin{subequations}
\begin{eqnarray}
E_1-E_0 = E_3-E_2 & = & \frac{1}{2} \left(
\sqrt{4J^2+(\Delta_1+\Delta_2)^2} -
\sqrt{4J^2+(\Delta_1-\Delta_2)^2} \right),
\\
E_2-E_0 = E_3-E_1 & = & \frac{1}{2} \left(
\sqrt{4J^2+(\Delta_1+\Delta_2)^2} +
\sqrt{4J^2+(\Delta_1-\Delta_2)^2} \right).
\end{eqnarray}
\end{subequations}
Note that this derivation shows that the selective darkening gate can also be used when \( \Delta_1 = \Delta_2 \), giving \( a_1 = a_2 \) for the required amplitude ratio for both qubit 1 or qubit 2 as the target qubit. In this special case the selective darkening method is equivalent to the method proposed by Beige et al. \cite{beige:00} and has recently been demonstrated experimentally \cite{filipp:11}.

\section{Relation to other gates}\label{sec:othergates}
Independently of our work on selective darkening (SD), a cross resonance (CR) scheme for realizing two-qubit gates for transversely-coupled qubits was proposed in reference \cite{rigetti:10}. The CR coupling scheme relies on driving one qubit (the control qubit) at the resonance frequency of the other qubit (the target qubit) and has the result of driving Rabi oscillations in the target qubit in one of two opposite directions based on the state of the control qubit. Even though at first sight the driving condition and resulting dynamics might seem to be very different from those of the SD scheme, the two schemes are related by a simple rotation. This relation can be seen by looking at figures 2(b) and 2(c) and noting that in both cases the two-qubit dynamics involve rotations of the target-qubit state about a single axis. As a result, the two-qubit evolution commutes with any single-qubit rotation about the same axis. One can therefore convert any CR-like operation into an SD-like operation by applying a single-qubit rotation designed to cancel one of the two possible CR rotations. Crucially, because the single-qubit and two-qubit rotations commute, they can be applied simultaneously. Starting from the CR scheme as a reference point, one could say that the driving field applied to the target qubit in the SD scheme is designed to oppose one of the two rotations in the CR scheme, thus producing CNOT-gate dynamics.

It is interesting in this context to look further back into the history of coupling schemes for superconducting qubits. After the proposal of the FLICFORQ coupling scheme \cite{rigetti:05}, it was shown that that this scheme is a special case of a family of coupling schemes using the physics of double resonance \cite{ashhab:06, ashhab:07}. In particular, even though the coupling scheme used in reference \cite{majer:07} might seem very different from FLICFORQ, they are both special cases of the double-resonance family of coupling schemes. The general condition that needs to be satisfied for double resonance is given by:

\begin{equation}
  \pm \sqrt{(\hbar\omega_1-\Delta_1)^2 + a_1^2}
  \pm \sqrt{(\hbar\omega_2-\Delta_2)^2 + a_2^2} =
  \hbar\omega_1 - \hbar\omega_2.
  \label{eq:doubleres}
\end{equation}

It is then interesting to note that the condition for the CR \cite{rigetti:10} and SD gates \cite{groot:10a}, i.e. driving both qubits at the frequency of one of them with small amplitudes, satisfies the double-resonance condition. One can then wonder whether these coupling schemes can also be seen as two more (closely-related) members of the double-resonance family.

The CR and SD schemes can indeed be seen as special cases of double resonance. However, they also involve some qualitative differences with previously studied cases, such as the FLICFORQ and ac-Stark-shift-induced resonance, making them more like distant relatives of the conventional double-resonance schemes (It is worth noting here that another different scheme of the same family was discovered in numerical results in reference \cite{ashhab:07}).
One difference between the more conventional double-resonance schemes and the CR and SD schemes becomes apparent if one plots the energy-level diagrams of the two driven qubits in the dressed-state picture. In the schemes of references~\cite{rigetti:05, majer:07}, one transition between dressed states of qubit 1 becomes resonant with one transition between dressed states of qubit 2.
As a result, a swap-like operation takes place with the two single-qubit transitions driven in opposite directions, such that the total energy is conserved. In the case of CR and SD, two pairs of transitions are resonant and are involved in the gate operation, as can be seen in figures 1 and 4 of reference \cite{rigetti:10}. The result turns out to be a conditional operation: the well-known CNOT-gate dynamics in the SD case, and a CNOT gate combined with a single-qubit rotation in the CR case.

The classification of different gates can also be done using the geometric representation of references~\cite{makhlin:02, zhang:03, geller:10}. Being related by simple single-qubit gates, the CR and SD CNOT gates correspond to exactly the same dynamics of the Makhlin parameters. The more conventional double-resonance class of gates, as well as the gate of reference~\cite{liu:06}, all exhibit oscillations corresponding to the iSWAP gate. As a result the Makhlin parameters exhibit similar behaviour for these cases.

In reference \cite{ferber:10} a scheme is discussed to control the individual states in degenerate subspaces of three qubits coupled in a loop. That scheme is based on the same principles as the SD and CR method. The gate proposed in reference \cite{liu:06} is similar to the CR and SD method in the sense that it also provides a microwave-only entangling gate between two qubits and does not involve shifting dressed states into resonance by strong driving. This gate however requires a coupling type that is at least partially longitudinal. Creating entangling evolution between two coupled qubits by tuning the amplitudes and phases of a driving field is also discussed in references \cite{li:08} and \cite{li:12}.

Another intimately related effect to SD is illustrated by the \emph{dark states} encountered in circuit quantum electrodynamics. When two qubits are both coupled to a harmonic oscillator, and the qubits are biased to have equal splitting, the energy levels exhibit an anti-crossing. Some of the transitions are typically found to be forbidden. This effect can be interpreted as SD, as the condition for selective darkening for two qubits of equal splitting is when both qubits are driven with equal amplitude \cite{filipp:11} (see section~\ref{sec:seldaB}). Recently, it has been proposed and demonstrated that these properties can be exploited further by defining a new logical qubit from two conventional qubits, using only two out of the four original states \cite{gambetta:11, srinivasan:11,hoffman:11}. Within this definition, the coupling to the resonator can be treated as tunable, and the qubit itself can be shielded from decay through the resonator.

\section{Time-domain evolution of the gate and spurious driving of other transitions}\label{sec:timedom}

  To compare different methods for creating entangling gates, or to design the optimum parameters for a certain experiment, it is important to know the potential sources of errors, and how these errors depend on the various parameters in the system. The first potential source of errors that we study is related to the ac-Stark effect, i.e. the energy shift that the driving field can induce on the non-resonant transitions in the system. Specifically, the applied driving field is resonant with the qubit that is to be flipped (the target qubit of the CNOT gate), but the same field also drives the control qubit, which typically has a different transition frequency. This off-resonant driving leads to a predictable, but unwanted deviation from the pure CNOT-type evolution of the system.

\subsection{Ideal gate evolution under the rotating wave approximation}\label{sec:timeideal}

The time evolution of the system is best studied in a suitable rotating frame. In this case the driving term \( \hat{H}_{\mathrm{drive}} \) in equation~(\ref{eq:Hdrive}) can be conveniently split into a (time-independent) co-rotating term \( \tilde{V} \) and a (time-dependent) counter-rotating term \( \tilde{W} \) (see appendix~\ref{apx:rotframe}).
For the co-rotating term \( \tilde{V} \) we find the matrix elements given in equation~(\ref{eq:transstr_strongJ}a-d) with \( \tilde{H}_{\mathrm{drive}} = \tilde{V} \).

We now make the rotating wave approximation: The
counter-rotating term $\tilde{W}$ oscillates very fast (at
frequency $2\omega$) and is therefore approximated by its average
value, zero. The gate we would like to perform is the usual CNOT gate,
with qubit 2 as the target qubit. We therefore
choose
\begin{subequations}
\begin{eqnarray}
\hbar\omega & = & E_3-E_2 = E_1-E_0 \nonumber
\\
& = & \frac{1}{2} \left( \sqrt{(\Delta_1+\Delta_2)^2+4J^2} -
\sqrt{(\Delta_1-\Delta_2)^2+4J^2} \right)
\label{eq:omega_driving_criteria}
\\
\frac{a_2}{a_1} & = &
\frac{\sin\theta_+}{\cos\theta_-}
\\
\varphi_2-\varphi_1 & = & 0.
\end{eqnarray}
\end{subequations}
We are now left with
\begin{eqnarray}
\tilde{H}_0 & = & \frac{1}{2} \sqrt{(\Delta_1-\Delta_2)^2+4J^2}\left( \begin{array}{cccc}
-1 & 0  & 0 & 0 \\
0  & -1 & 0 & 0 \\
0  & 0  & 1 & 0 \\
0  & 0  & 0 & 1 \\
\end{array}
\right)
\\
\bra{1} \tilde{V} \ket{0} & = & 0 \nonumber
\\
\bra{2} \tilde{V} \ket{0} & = & \frac{a_1}{2} \left(
\cos\theta_+ + \sin\theta_+ \tan\theta_- \right) e^{-i\varphi_1} \nonumber
\\
\bra{3} \tilde{V} \ket{2} & = & a_1
\sin\theta_+ e^{-i\varphi_1} \nonumber
\\
\bra{3} \tilde{V} \ket{1} & = & \frac{a_1}{2} \left(
\cos\theta_+ - \sin\theta_+
\tan\theta_- \right) e^{-i\varphi_1} \nonumber
\\
\bra{i} \tilde{V} \ket{j} & = & \bra{j} \tilde{V} \ket{i}^*.
\end{eqnarray}

In a first approximation, we ignore the matrix element
$\tilde{V}_{20}$ and $\tilde{V}_{31}$ (and their Hermitian
conjugates). We then find the effective Hamiltonian
\begin{equation}
  \tilde{H} =
\tilde{H}_0 + \left( \begin{array}{cccc}
0 & 0 & 0 & 0 \\
0 & 0 & 0 & 0 \\
0 & 0 & 0 & \tilde{V}_{32}^* \\
0 & 0 & \tilde{V}_{32} & 0 \\
\end{array}
\right).
\end{equation}
These two terms commute, and their effects can be easily superimposed.
If the Hamiltonian is allowed to operate for a time
$t=h/(4|\tilde{V}_{32}|)=h/(4a_1\sin\theta_+)$,
the first term produces the transformation
\begin{equation}
\left( \begin{array}{cccc}
e^{i\sqrt{(\Delta_1-\Delta_2)^2+4J^2}t/2} & 0
& 0 & 0 \\
0 & e^{i\sqrt{(\Delta_1-\Delta_2)^2+4J^2}t/2} & 0
& 0 \\
0 & 0 & e^{-i\sqrt{(\Delta_1-\Delta_2)^2+4J^2}t/2} & 0 \\
0 & 0 & 0 & e^{-i\sqrt{(\Delta_1-\Delta_2)^2+4J^2}t/2} \\
\end{array}
\right),
\end{equation}
and the second term produces the transformation
\begin{equation}
\left( \begin{array}{cccc}
1 & 0 & 0 & 0 \\
0 & 1 & 0 & 0 \\
0 & 0 & 0 & -i e^{-i\varphi_1} \\
0 & 0 & -i e^{i\varphi_1} & 0 \\
\end{array}
\right).
\end{equation}
These transformations are given in the rotating frame. They can be
transformed back to the lab frame to give:
\begin{equation}
\left( \begin{array}{cccc}
e^{-iE_0 t} & 0 & 0 & 0 \\
0 & e^{-iE_1 t} & 0 & 0 \\
0 & 0 & e^{-iE_2 t} & 0 \\
0 & 0 & 0 & e^{-iE_3 t} \\
\end{array}
\right) \left( \begin{array}{cccc}
1 & 0 & 0 & 0 \\
0 & 1 & 0 & 0 \\
0 & 0 & 0 & -i e^{-i\varphi_1} \\
0 & 0 & -i e^{i\varphi_1} & 0 \\
\end{array}
\right).
\label{eq:iCNOT}
\end{equation}
The leftmost matrix represents the Larmor precession of the system in the lab frame. Taking \( \varphi_1 = 0 \), the rightmost matrix is exactly the CNOT gate, up to a single-qubit phase gate and a global phase factor (meaning that the Makhlin parameters~\cite{makhlin:02} are equal to those of the standard CNOT gate).

\subsection{Single- and two-qubit errors caused by the ac-Stark effect} \label{sec:acStark}

We now include the two (or four including the conjugate terms) matrix elements that we have ignored in the previous subsection:
\begin{subequations}
\begin{eqnarray}
\bra{2} \tilde{V} \ket{0} & = & \frac{a_1}{2} \left(
\cos\theta_+ + \sin\theta_+
\tan\theta_- \right) e^{-i\varphi_1}
\\
\bra{3} \tilde{V} \ket{1} & = & \frac{a_1}{2} \left(
\cos\theta_+ - \sin\theta_+
\tan\theta_- \right) e^{-i\varphi_1}.
\end{eqnarray}
\end{subequations}
These matrix elements produce an ac-Stark shift on qubit 1 (the control qubit). If
 $\theta_+$ and $\theta_-$ are small, the above two matrix
elements are approximately equal to $a_1/2$ (up to a phase
factor). This matrix element then causes a shift of
$a_1^2/2(E_2-E_0-\hbar\omega)$. This frequency shift can be
incorporated easily into the transformation given in equations~(\ref{eq:phitilde}-\ref{eq:VWtilde}),
simply modifying the energies that enter in equation~(\ref{eq:Htimedom}).

Note however that the two
matrix elements above are slightly different, so that the ac-Stark
shift of qubit 1 depends on the state of qubit 2. This situation
would induce an unwanted controlled-phase (CPHASE) gate. If we take
$\Delta_1/h=6$~GHz, $\Delta_2/h=5$~GHz and $J/h=0.1$~GHz, we find
\( \theta_1 = 0.0058 \pi \) and \( \theta_2 = 0.063 \pi \),
which gives the ratio between the matrix elements:
\begin{equation}
\frac{\bra{2} \tilde{V} \ket{0}}{\bra{3} \tilde{V} \ket{1}} =
\frac{\cos\theta_+ + \sin\theta_+ \tan\theta_-}
     {\cos\theta_+ - \sin\theta_+ \tan\theta_-} = 1.02.
\end{equation}
This means that for the given parameters, the two different values
for the ac-Stark shift
will differ by about 4\%. This 4\% difference of the frequency shift should
therefore be designed such that it causes an overall phase shift
much smaller than $2\pi$ over the gate time of typically a few nanoseconds.
It can then be neglected. Otherwise, it could harm the fidelity of the
gate. Even for extremely strong driving ($a_1/h=$1 GHz), which gives a large
ac-Stark shift (about 0.5 GHz), over a gate time of 4 ns, we
obtain an unwanted CPHASE rotation by an angle of $0.16\pi$. For weaker
driving this unwanted angle will be smaller.

Note that in our case, this CPHASE-type error can also be corrected relatively easily using single-qubit gates. In this case the SD-CNOT part of the gate is made intentionally to deviate slightly from a perfect CNOT gate by using a longer or shorter gate time, such that the evolution of the state-vector overshoots or undershoots the normal \( \pi \)-rotation. If the correction to the gate time is chosen appropriately, the combined evolution of the unwanted CPHASE gate and the intentionally driven CNOT gate can be transformed into a perfect CNOT gate using single-qubit gates before and after the two-qubit driving. This effect is related to the well-known fact that a CPHASE gate can be transformed into a CNOT gate using only single-qubit gates \cite{nielsen:00}. This type of correction will be demonstrated in section~\ref{sec:num2level}.

A more rigorous analysis (see appendix~\ref{apx:maxspeed}), where the driving field is treated quantum mechanically, confirms the single-qubit and two-qubit errors found here. In addition, even though we mostly look at the high driving limit where the classical description of the field gives accurate results, some phenomena can be identified more easily. Specifically it is shown that for increasing amplitudes of the driving field not only the levels of the control qubit are shifted, but also the transitions of the target qubit are no longer perfectly degenerate. In this case no transition can be fully darkened. The result is a single-qubit-type error on the target qubit. As mentioned above, single-qubit errors can be corrected relatively easily. It should be noted however that this error can slow down the CNOT gate. A perfect SD-CNOT gate requires the state vectors in the example of figure~\ref{fig:bloch} to be on opposite sides of the Bloch sphere at the end of the gate, and co-rotation of the transition that should be darkened leads to a saturation of the effective gate speed. This will also be considered in section~\ref{sec:num2level}.

\section{Numerical simulation of gate dynamics for two-level qubits}\label{sec:num2level}

To quantify the fidelity of the selective darkening CNOT (SD-CNOT) gate, the evolution of
the system was simulated numerically. We take the driving conditions derived from the simple weak driving case given in section~\ref{sec:selda}: the frequency of the driving field was chosen to be \( \omega = (E_1-E_0)/\hbar = (E_3-E_2)/\hbar \), and the amplitude ratio as given in equation~(\ref{eq:sd_condition_strongdr}). The amplitudes are ramped up and down using the half-sine envelope function \( \sin(\pi t/t_{\mathrm{gate}}) \), with \( t \) running from 0 to the total gate time \( t_{\mathrm{gate}} \). Note that we chose to gradually turn on and off the driving field rather than using step functions, because in many cases the steep edges can lead to undesired excitations in the system. This approach increases the gate time with a factor of \( \pi/2 \) compared to the minimum gate time where the driving amplitude is turned on and off abruptly. Other envelope shapes would give similar correction factors.

For a given maximum amplitude \( a_1 \), the evolution of the system was simulated. The evolution operator at the end of the gate was evaluated according to its overlap with the ideal CNOT gate, which is quantified using the fidelity \( \mathcal{F} = \mathrm{Tr}( M_{\mathrm{gate}} M_{\mathrm{CNOT}})/4 \). Here \( M_{\mathrm{gate}} \) is the (unitary) matrix of the simulated gate and \( M_{\mathrm{CNOT}} \) the ideal CNOT gate from equation~(\ref{eq:iCNOT}) (which includes the phase factors \( i \)). This calculation is performed for 20 different gate times around an estimated optimum. An interpolating polynomial function was used to fit the dependence of the fidelity on the gate time, providing the final accurate estimate for the optimum gate time. The gate was recalculated using this gate time, giving the final result for the fidelity. This sequence is repeated for different driving amplitudes \( a_1 \) (with \( a_2 \) adjusted to the ratio given in equation~(\ref{eq:sd_condition_strongdr})). The used qubit splittings are \( \Delta_1/h = 6~\mathrm{GHz} \) and \( \Delta_2/h = 5~\mathrm{GHz} \) and the phases of the driving fields are \( \varphi_1 = \varphi_2=0 \).

\begin{figure}[htbp]
  \begin{center}
    \includegraphics{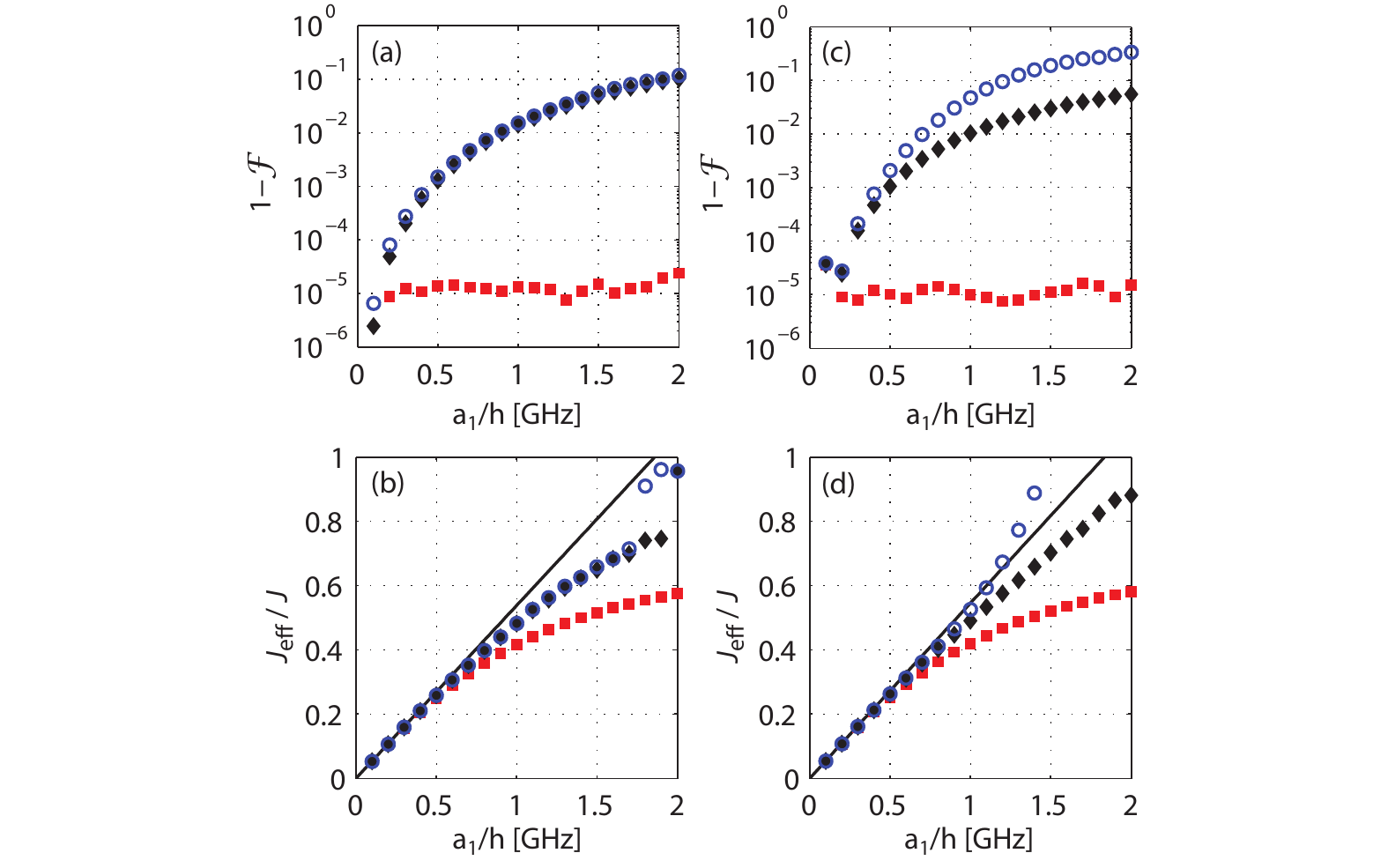}
  \end{center}
  \caption{Numerical simulation of a SD-CNOT gate for 2-level qubits. Panels (a) and (b) are calculated for coupling strengths of \( J/h=100~\mathrm{MHz} \) and (c),(d) for \( J/h=25~\mathrm{MHz} \).  Blue circles represent the results for the SD-CNOT gate, incorporating only corrections for Larmor precession and ac-Stark shift of the control qubit. The other two data sets represent the SD-CNOT gate including error corrections using single-qubit phase gates (black diamonds) and arbitrary single-qubit gates (red squares). In panels (a) and (c) the gate error  \(1-\mathcal{F}\) is plotted versus the driving strength. Panels (b) and (d) give the corresponding gate speed, expressed as the effective coupling strength \( J_{\mathrm{eff}} \) divided by the natural coupling strength, for which the fidelity \(\mathcal{F}\) is achieved. The black lines indicate the expected dependence from equation~(\ref{eq:transstr_strongJ}). The values used for the qubit splittings are \( \Delta_1/h = 6~\mathrm{GHz} \) and \( \Delta_2/h = 5~\mathrm{GHz} \).}
  \label{fig:2level}
\end{figure}

The results of the simulations are shown in figure~\ref{fig:2level}. Panels (a) and (b) were calculated for a coupling strength \( J/h = 100~\mathrm{MHz}\), and panels (c) and (d) for \( J/h = 25~\mathrm{MHz} \).
The blue circles give the result for the procedure as described above, using the lowest number of corrections. We only take into account the Larmor precession, and for the control qubit the ac-Stark and Bloch-Siegert shift due to the (off-resonant) driving, disregarding for the level shifts any effects related to the coupling. The bare SD-CNOT gate is calculated first, and the corrections are applied afterward as phase gates. The required phase for the ac-Stark and Bloch-Siegert shift corrections is calculated analytically from the energy shifts \( \delta E_{\rm ac-Stark} = a_1^2/2(E_2-E_0-\hbar\omega) \) and $\delta E_{\rm Bloch-Siegert} = a_1^2/2 (E_2-E_0+\hbar\omega)$ and the gate time \( t_{\mathrm{gate}} \), taking into account the envelope shape. The correction for the Larmor precession is given by the conjugate of the leftmost matrix in equation~(\ref{eq:iCNOT}). Note however that all of these corrections often do not have to be implemented physically in an experiment. The internal phase evolution of a microwave-source that is set to the resonance frequency will automatically follow the Larmor precession of the qubit, and a single-qubit phase gate can often be incorporated easily in preceding or following single-qubit gates.

For the black diamonds and red squares we performed the same procedure, but in addition we allowed single-qubit phase gates (black diamonds) or arbitrary single-qubit rotations (red squares) before and after the two-qubit gate. For each simulation of the two-qubit gate an optimization algorithm is used to find single-qubit gates that optimize the fidelity of the gate. This approach allows for correcting certain error types, such as the CPHASE-type error discussed in section~{\ref{sec:acStark}}. Note that in all the three cases, the corrections influence not only the maximum fidelity of the gate, but also for which \( t_{\mathrm{gate}} \) the maximum is found.

We characterized the gate speed as the effective coupling strength \( J_{\mathrm{eff}} = h/t_{\mathrm{gate}}\cdot\pi/2 \), where the factor \( \pi/2 \) takes into account the envelope function of the driving field. In panels (b) and (d), the black lines represent the expected gate speed as given in equation~(\ref{eq:transstr_strongJ}).
For low driving strengths the gate speed follows exactly the linear behaviour as expected from the transition strength. The gate error (given by \(1-\mathcal{F}\)), increases monotonically with driving strength, except for the case with arbitrary single-qubit corrections where the error is \( \sim 10^{-5} \) throughout the whole range and is limited only by the accuracy of the numerical calculation.
The low error in the latter case shows that the errors in the SD-CNOT gate are either single-qubit errors, or two-qubit errors that can be incorporated in the combined evolution to form a perfect CNOT gate, such as the CPHASE-type errors discussed in section~\ref{sec:acStark}.
Also in experiment this would be an attractive way to enhance the gate performance.

The slow-down of the gate for the results with arbitrary single-qubit gates (red squares) at higher driving frequencies can be explained with the error types discussed in the previous sections. If the transition \( \ket{0} \leftrightarrow \ket{1} \) is not fully darkened, and the state vector rotates in the same direction as the \( \ket{2} \leftrightarrow \ket{3} \) (CNOT) transition, a longer gate time is required to achieve the maximum separation of the two state vectors (to achieve a perfect CNOT gate the alignment of the two state vectors should be exactly opposite). Alternatively, a CPHASE-type error can require a slight overshoot of the CNOT gate evolution to achieve the best fidelity when including the single-qubit gates. Both processes are likely to contribute to the observed slow-down of the gate.

At higher driving strengths the gate speed starts to deviate from the linear dependence of the weak-driving limit. Especially where the error is larger than 0.1, the gate fidelity becomes less well-behaved. For \( J=25~\mathrm{MHz} \), the effective coupling strength even seems to exceed the natural coupling strength, but this can be explained simply by the fact that the performed gate is no longer close to the desired gate. Also for \( J=25~\mathrm{MHz} \), we see an improvement when using the single-qubit phase corrections (black diamonds).

Even without the extra gate optimizations (blue circles), we find a high fidelity at relatively high gate speeds. For the coupling strength of \( J=100~\mathrm{MHz} \), gate speeds corresponding to \( 0.48J \) and \( 0.07J \) give a gate fidelity of 99\% and 99.99\% respectively. The exact numbers will depend on the system parameters. Note that these results were obtained using a fixed driving frequency and amplitude ratio, obtained from simple analytic equations, and using a fixed pulse shape. Using more advanced optimization schemes, for example using the DRAG (Derivative Removal by Adiabetic Gate) method \cite{tian:00, motzoi:09}, these number can likely be enhanced further.

\section{Weakly anharmonic qubits}\label{sec:weakanharmonic}
  The above analysis of the SD-CNOT gate and the possible errors is accurate for true two-level qubits. However, many qubit implementations are in fact multi-level systems. In this case a nonlinear level structure allows the system to be used as an \emph{effective} two-level system, where only two out of a manifold of levels are used as the (computational) qubit states.
When dealing with weakly anharmonic qubits, i.e., qubits where the transition
frequencies between consecutive energy levels differ only slightly, the additional energy levels have to be taken into account, as they can significantly influence the dynamics of the system. For example, one obvious complication
that arises in this case is the leakage to these additional states.

We take the computational basis to be formed by the lowest two energy states of the qubits, labeled  $\ket{00}$, $\ket{01}$, $\ket{10}$ and $\ket{11}$, and the states outside the computational basis are labeled $\ket{02}$, $\ket{20}$, etc. Taking into account the third level of each qubit, and
making the approximation of weak coupling
($J\ll\Delta_1-\Delta_2$) and weak anharmonicity ($\delta_j \equiv
\hbar\omega_{1\leftrightarrow 2;j} - \hbar\omega_{0\leftrightarrow 1;j} \ll
\Delta_1 - \Delta_2$; note that under this definition $\delta_j$ is negative for phase
and transmon qubits and positive for capacitively shunted flux qubits), we find the following approximate
expressions for the energy eigenstates:
\begin{eqnarray}
\ket{0} & = & \ket{00} \nonumber
\\
\ket{1} & = & \ket{01} - \frac{J}{\Delta_1-\Delta_2} \ket{10} \nonumber
\\
\ket{2} & = & \ket{10} + \frac{J}{\Delta_1-\Delta_2} \ket{01} \nonumber
\\
\ket{3} & = & \ket{11} +
\frac{\sqrt{2}J}{\Delta_1-\Delta_2-\delta_2} \ket{02} -
\frac{\sqrt{2}J}{\Delta_1-\Delta_2+\delta_1} \ket{20} \nonumber
\\
\ket{4} & = & \ket{02} -
\frac{\sqrt{2}J}{\Delta_1-\Delta_2-\delta_2} \ket{11} \nonumber
\\
\ket{5} & = & \ket{20} +
\frac{\sqrt{2}J}{\Delta_1-\Delta_2+\delta_1} \ket{11} \nonumber
\\
\ket{6} & = & \ket{12} +
\frac{2J}{\Delta_1-\Delta_2-\delta_1-\delta_2} \ket{21} \nonumber
\\
\ket{7} & = & \ket{21} +
\frac{2J}{\Delta_1-\Delta_2+\delta_1-\delta_2} \ket{12} \nonumber
\\
\ket{8} & = & \ket{22}.
\end{eqnarray}
The respective approximate energies are given by:
\begin{eqnarray}
E_0 & = & - \frac{\Delta_1}{2} - \frac{\Delta_2}{2} \nonumber
\\
E_1 & = & - \frac{\Delta_1}{2} + \frac{\Delta_2}{2} -
\frac{J^2}{\Delta_1-\Delta_2} \nonumber
\\
E_2 & = & + \frac{\Delta_1}{2} - \frac{\Delta_2}{2} +
\frac{J^2}{\Delta_1-\Delta_2} \nonumber
\\
E_3 & = & + \frac{\Delta_1}{2} + \frac{\Delta_2}{2} +
\frac{2J^2}{\Delta_1-\Delta_2-\delta_2} -
\frac{2J^2}{\Delta_1-\Delta_2+\delta_1} \nonumber
\\
E_4 & = & - \frac{\Delta_1}{2} + \frac{3\Delta_2}{2} + \delta_2 -
\frac{2J^2}{\Delta_1-\Delta_2-\delta_2} \nonumber
\\
E_5 & = & + \frac{3\Delta_1}{2} - \frac{\Delta_2}{2} + \delta_1 +
\frac{2J^2}{\Delta_1-\Delta_2+\delta_1} \nonumber
\\
E_6 & = & + \frac{\Delta_1}{2} + \frac{3\Delta_2}{2} + \delta_2 -
\frac{4J^2}{\Delta_1-\Delta_2+\delta_1-\delta_2} \nonumber
\\
E_7 & = & + \frac{3\Delta_1}{2} + \frac{\Delta_2}{2} + \delta_1 +
\frac{4J^2}{\Delta_1-\Delta_2+\delta_1-\delta_2} \nonumber
\\
E_8 & = & + \frac{3\Delta_1}{2} + \frac{3\Delta_2}{2}.
\end{eqnarray}
Note that the labels are ordered for increasing energy of the states, except for the interchange of labels 3 and 4. This interchange makes that states 0-3 correspond to the two-level qubit case, and makes it easier to compare the two cases.

Let us consider the CNOT gate with qubit 2 being the target qubit.
In the qubit basis, this means that the $\ket{0}\rightarrow\ket{1}$
transition is to be darkened, and the $\ket{2}\rightarrow\ket{3}$
transition provides the CNOT gate dynamics.
The required relation between the driving amplitudes is again given
by equation~(\ref{eq:AmplitudeRatio})
and $\varphi_2-\varphi_1=0$. For simplicity in the expressions
we also take \( \varphi_1 = \varphi_2 = 0 \). The transition strengths of the darkened and CNOT-gate transition are now given by
\begin{subequations}
\begin{eqnarray}
  \bra{1} \frac{a_1}{2} \hat{\sigma}_x^{(1)} + \frac{a_2}{2} \hat{\sigma}_x^{(2)} \ket{0}
& \approx & \frac{a_1}{2} \left( - \frac{J}{\Delta_1-\Delta_2} +
\frac{J}{\Delta_1-\Delta_2} \right) = 0
\\
\bra{3} \frac{a_1}{2} \hat{\sigma}_x^{(1)} + \frac{a_2}{2} \hat{\sigma}_x^{(2)} \ket{2}
& \approx & \frac{a_1}{2} \left( \left[ - \frac{2J}{\Delta_1-\Delta_2+\delta_1} +
\frac{J}{\Delta_1-\Delta_2} \right] + \frac{J}{\Delta_1-\Delta_2}
\left[ 1 + \frac{2J^2}{(\Delta_1-\Delta_2)^2} \right] \right) \nonumber
\\
& \approx & a_1 \frac{J}{\Delta_1-\Delta_2}
\left( \frac{\delta_1}{\Delta_1-\Delta_2} +
\frac{J^2}{(\Delta_1-\Delta_2)^2} \right),
\label{eq:Tsmalllin}
\end{eqnarray}
\end{subequations}
where \( \hat{\sigma}_{x}^{(1)} = \hat{\sigma}_{x} \otimes \mathbbm{1} \) and \( \hat{\sigma}_{x}^{(2)} = \mathbbm{1} \otimes \hat{\sigma}_{x} \) are redefined using
\begin{equation}
  \hat{\sigma}_x =
  \left( \begin{array}{ccc}
  0 & 1        & 0       \\
  1 & 0        & \sqrt 2 \\
  0 & \sqrt 2 & 0        \\
  \end{array}
  \right).
  \label{eq:sigmax3}
\end{equation}
From these equations we can recognise the first consequence of the additional levels: The transition strength of the \( \ket{3} \leftrightarrow \ket{2} \) (CNOT) transition is reduced by a potentially significant factor. This reduction is a consequence of the mixing between the states $\ket{11}$ and
$\ket{20}$, which introduces a new term into the matrix element
$\bra{3} \frac{a_1}{2} \hat{\sigma}_x^{(1)} + \frac{a_2}{2} \hat{\sigma}_x^{(2)} \ket{2}$.
For systems where both the anharmonicity \emph{and} the coupling energy are small compared to the detuning between the qubits, the speed of the gate will be greatly suppressed. So in order to achieve large gate speeds with weakly anharmonic qubits, one must design the system with one or both of these two factors relatively large.

Next we analyze the leakage to the states outside the computational basis.
The largest leakage arises from the transitions $\ket{1}\rightarrow\ket{4}$ and $\ket{3}\rightarrow\ket{6}$, because qubit 1 (the control qubit) is driven the strongest.
The corresponding matrix elements and energy detunings are given by:
\begin{subequations}
\begin{eqnarray}
\bra{4} \frac{a_1}{2} \hat{\sigma}_x^{(1)} + \frac{a_2}{2} \hat{\sigma}_x^{(2)} \ket{1}
& \approx & a_1 \frac{J}{\Delta_1-\Delta_2} \frac{1}{\sqrt{2}}
\left( -\frac{\delta_2}{(\Delta_1-\Delta_2)} +
\frac{J^2}{(\Delta_1-\Delta_2)^2} \right)
\\
\bra{6} \frac{a_1}{2} \hat{\sigma}_x^{(1)} + \frac{a_2}{2} \hat{\sigma}_x^{(2)} \ket{3}
& \approx & a_1 \frac{J}{\Delta_1-\Delta_2}
(1+\frac{\sqrt{2}}{4})
\end{eqnarray}
\end{subequations}
and
\begin{subequations}
\begin{eqnarray}
(E_4-E_1)-(E_1-E_0) & \approx & \delta_2 -
\frac{2J^2\delta_2}{(\Delta_1-\Delta_2)^2}
\\
(E_6-E_3)-(E_3-E_2) & = & \delta_2 -
\frac{3J^2}{\Delta_1-\Delta_2}.
\end{eqnarray}
\end{subequations}
To avoid leakage, the transition strengths of the unwanted transitions should
be lower (possibly much lower) than the detuning with the driving fields (similar as in section~\ref{sec:seldaA}): \( \bra{l} \tilde{H}_{\rm drive} \ket{k} < |\hbar\omega - \hbar\omega_{l \leftrightarrow k}| \). Both leakage transitions have approximately equal energy splitting (\( \approx \delta_2 \)). The \( \ket{3} \leftrightarrow \ket{6}\) transition clearly has the highest transition strength. Using this transition we find the estimate for the maximum allowed Rabi frequency of the CNOT transition:
\begin{equation}
  \omega_{R, {\rm max}} = \frac{\delta_2}{2+\sqrt{2}/2}
  \left[\frac{\delta_1}{(\Delta_1-\Delta_2)} + \frac{J^2}{(\Delta_1-\Delta_2)^2} \right].
  \label{eq:wRmax}
\end{equation}
How close one can get to this estimated maximum is simulated
numerically, and will be discussed shortly.

A third issue that one needs to keep in mind when dealing with
weakly anharmonic qubits is that the additional energy
levels cause frequency shifts that break the symmetry
$E_3-E_2=E_1-E_0$. In particular, the energies of the
state 4 (\( \approx \ket{02} \)) and 5 (\( \approx \ket{20} \)) will be close to the energy of the
state 3 (\( \approx \ket{11} \)). Fortunately, they cause shifts in opposite
directions and almost cancel. The net shift is given by (see the
expressions for the energies above):
\begin{equation}
\delta E_3 = \frac{2J^2}{\Delta_1-\Delta_2-\delta_2} - \frac{2J^2}{\Delta_1-\Delta_2+\delta_1}
\approx \frac{2J^2(\delta_1+\delta_2)}{(\Delta_1-\Delta_2)^2}.
\end{equation}
This frequency shift is not specific for the selective darkening  method, but typical for
weakly anharmonic qubits, and results in an always-on interaction that
drives a CPHASE gate.

\begin{figure}[htbp]
  \begin{center}
    \includegraphics{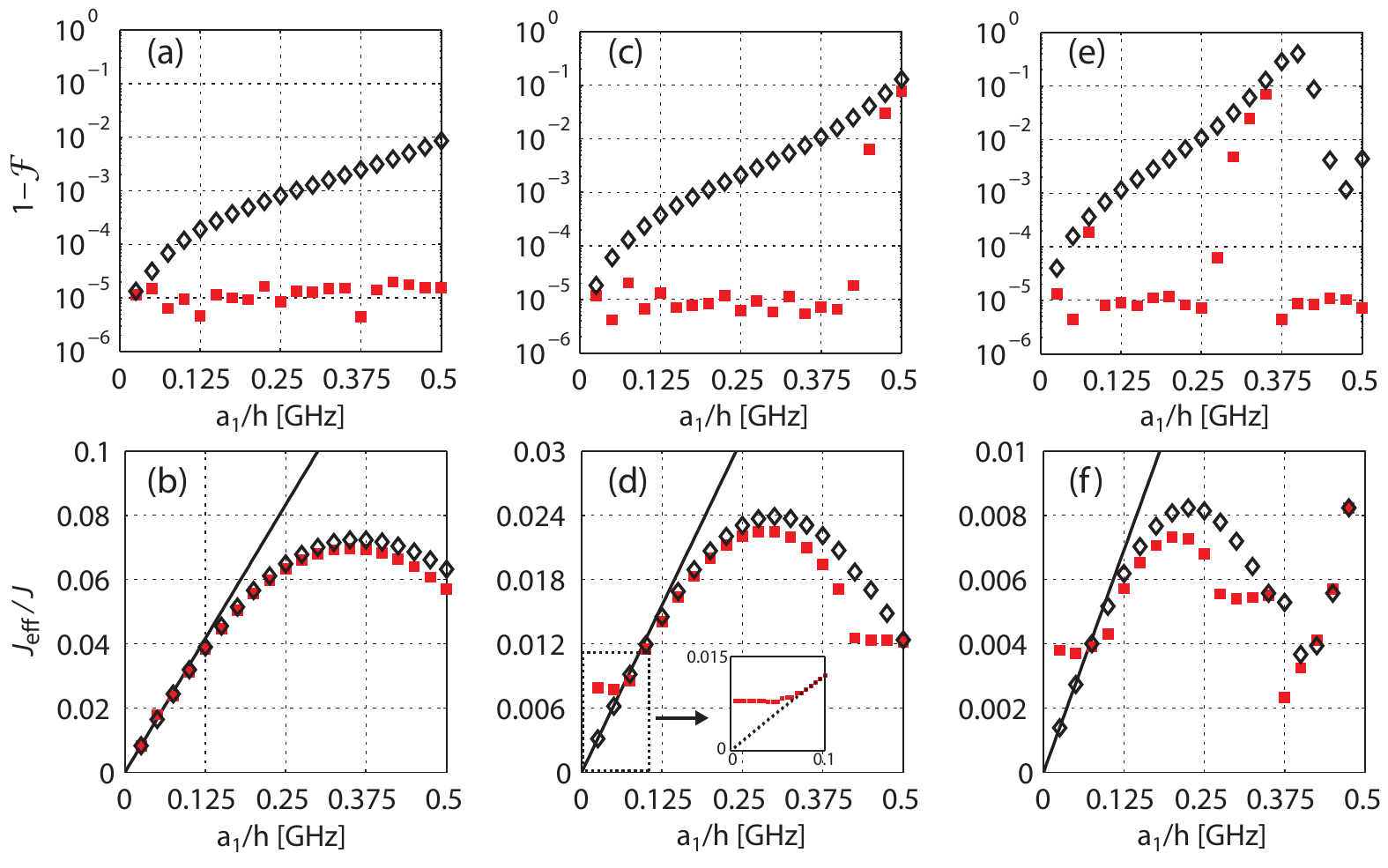}
  \end{center}
  \caption{Numerical simulation of a SD-CNOT gate for 3-level qubits with weak anharmonicity. Black diamonds represent the SD-CNOT gate including single-qubit phase gate corrections and the red squares represent the SD-CNOT gate including arbitrary single-qubit gates.
  Panels (a,b) are calculated for anharmonicities of \( \delta_1/h=-400~\mathrm{MHz} \),  \( \delta_2/h=-360~\mathrm{MHz} \), (c,d) for \( \delta_1/h=-200~\mathrm{MHz} \),  \( \delta_2/h=-180~\mathrm{MHz} \) and (e,f) for \( \delta_1/h=-100~\mathrm{MHz} \),  \( \delta_2/h=-90~\mathrm{MHz} \).
  In panels (a,c,e) the gate error \(1-\mathcal{F}\) is plotted versus the driving strength \(a_1\). Panels (b,d,f) give the corresponding gate speed \( J_{ \mathrm{eff} }/J\) for which the fidelity is achieved. Dashed lines indicate the expected dependence from equation~(\ref{eq:Tsmalllin}). The qubit parameters used are \( \Delta_1/h = 6~\mathrm{GHz} \), \( \Delta_2/h = 5~\mathrm{GHz} \) and \( J/h=25~\mathrm{MHz} \).}
  \label{fig:3level}
\end{figure}

We again perform numerical simulations to determine the total fidelity of all
the combined errors. First we start with simulations where we vary the anharmonicity in order to see its effect on the gate speed and fidelity. We use three different combinations of anharmonicity: (\( \delta_1/h=-400~\mathrm{MHz} \),  \( \delta_2/h=-360~\mathrm{MHz} \)), (\( \delta_1/h=-200~\mathrm{MHz} \),  \( \delta_2/h=-180~\mathrm{MHz} \)) and (\( \delta_1/h=-100~\mathrm{MHz} \),  \( \delta_2/h=-90~\mathrm{MHz} \)).
The other qubit parameters are \( \Delta_1/h=6~\mathrm{GHz} \), \( \Delta_2/h=5~\mathrm{GHz} \) and \( J/h=25~\mathrm{MHz} \).

The results are shown in figure~\ref{fig:3level}. At low driving strength, the gate speeds have a good correspondence with the expected dependence from the transition strength [equation~(\ref{eq:Tsmalllin})], confirming the slowdown of the SD-CNOT gate for decreasing anharmonicity. This implies that for weakly anharmonic qubits it is recommended to design the detuning between the qubits to be relatively small, so the negative impact on the gate speed due to the weak anharmonicity can be minimized.
Due to the low gate speeds, all the energy shifts induced by the driving and the higher-lying levels are also more significant. For this reason, the simple analytical calculation of the AC-Stark and Bloch-Siegert shifts of the control qubit (as was used for the calculation of the data represented by the blue circles in figure~\ref{fig:2level}) no longer gives useful results. Therefore we do not plot the corresponding results in figure~\ref{fig:3level}.

Another prominent feature is the decrease of the gate speed at high driving amplitudes. One possible explanation for this effect is a competition between the SD-CNOT gate and an AC-Stark induced CPHASE gate.
For figure~\ref{fig:3level}(d) a plateau is observed at low driving strength. This plateau is related to the CPHASE evolution caused by level shifts induced by the higher energy levels of the qubits, and is therefore independent of the driving amplitude. As mentioned before, a CPHASE-type error can be corrected by altering the duration of the two-qubit driving and using single-qubit rotations. The plateau indicates that the CPHASE contribution to the final CNOT gate is even dominating over the selective-darkening contribution.
The origin of the plateaus in (d) and (f) at high driving power is not known. The same holds for the sharp sudden increase in the error. Note however that at these points the driving strength is already higher than would be advisable given the anharmonicity of the qubits.

A common approach to improve gate speed and fidelity for weakly anharmonic qubits
is to use pulse shaping (see for example the DRAG method \cite{tian:00, motzoi:09}).
This approach could lead to improved gate speed and fidelity, but will not be discussed here.

\begin{figure}[htbp]
  \begin{center}
    \includegraphics{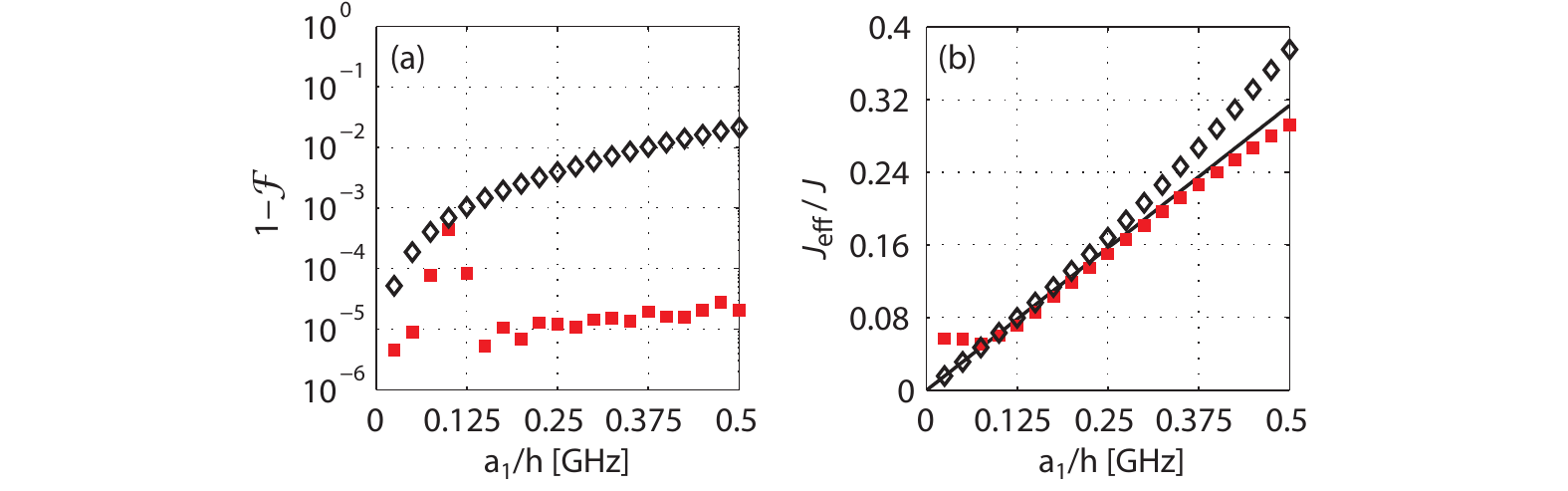}
  \end{center}
  \caption{Numerical simulation of a SD-CNOT gate for 3-level qubits with parameters similar to the experiment in reference \cite{chow:11}. The qubit parameters used are \( \Delta_1/h = 5.854~\mathrm{GHz} \), \( \Delta_2/h = 5.528~\mathrm{GHz} \), \( \delta_1/h=225~\mathrm{MHz} \),  \( \delta_2/h=255~\mathrm{MHz} \) and \( J/h=6~\mathrm{MHz} \). Black diamonds represent the SD-CNOT gate including single-qubit phase gate corrections and the red squares represent the SD-CNOT gate including arbitrary single-qubit gates.
  In panel (a) the gate error \(1-\mathcal{F}\) is plotted versus the driving strength \(a_1\). Panel (b) gives the corresponding gate speed \( J_{ \mathrm{eff} }/J\) for which the fidelity is achieved. The black line indicates the expected dependence from equation~(\ref{eq:Tsmalllin}). }
  \label{fig:ibm}
\end{figure}

In the experiment described in \cite{chow:11}, a cross-resonance (CR) gate is performed. Given the close relationship between the SD and CR method, we calculate, for comparison, the SD performance for the described system. The qubit parameters in this case are \( \Delta_1/h = 5.854~\mathrm{GHz} \), \( \Delta_2/h = 5.528~\mathrm{GHz} \), \( \delta_1/h=225~\mathrm{MHz} \),  \( \delta_2/h=255~\mathrm{MHz} \) and \( J/h=6~\mathrm{MHz} \). The coupling \( J \) is estimated from the numbers provided in the reference. Figure~\ref{fig:ibm} shows the result of the simulations. The gate speed in the simulations is in good agreement with the typical gate speeds in the experiment. With these system parameters one achieves a relatively high effective coupling, primarily thanks to the small detuning between the qubits. At very low driving strength we see a similar plateau as in the previous parameter set, and again attribute this to level shifts due to the states outside the computational basis. From the fidelity data in panel (a), we conclude that the leakage to these states is negligible at these driving strengths: if there were significant leakage to these levels, the results using the arbitrary single-qubit corrections represented by the red squares could not lead to the gate improvement that is observed, because the single-qubit corrections that we apply in the calculations cannot transfer back any state-population that is lost during the SD gate. The increase of the gate error for the results with arbitrary single-qubit gate corrections at \( a_1/h=0.1~\mathrm{GHz} \) is not yet understood. Also we do not see the significant slow-down of the gate at high driving strength as was measured in the experiment. A possible explanation for this is that different pulse-shapes were used, that in the experiment also DRAG-type \cite{tian:00, motzoi:09} pulses were applied, or that the presence of the resonator has an additional influence that is not covered by our direct-coupling model. Nevertheless, the calculation illustrates that similar performance can be expected for the SD and CR methods.

\section{Scalability}\label{sec:scalability}
Let us now consider the scalability of the SD-CNOT gate to systems of more than two qubits. We take a system
composed of a one-dimensional chain of $N$ transversely-coupled qubits and use the
Jordan-Wigner (JW) transformation to turn it into a system of
fermions occupying a one-dimensional lattice with $N$ sites. The
transformation is given by:
\begin{subequations}
\begin{eqnarray}
\hat{\sigma}_+^{(i)} & = & \prod_{j<i} \left( 1 - 2
\hat{c}_j^{\dagger} \hat{c}_j \right) \hat{c}_i \\
\hat{\sigma}_-^{(i)} & = & \prod_{j<i} \left( 1 - 2
\hat{c}_j^{\dagger} \hat{c}_j \right) \hat{c}_i^{\dagger} \\
\hat{\sigma}_z^{(i)} & = &  1 - 2 \hat{c}_i^{\dagger} \hat{c}_i,
\end{eqnarray}
\end{subequations}
where \( \hat{c} \) and \( \hat{c}^\dagger \) are annihilation and creator operators,
\begin{equation}
  \hat{\sigma}_+ = \left( \begin{array}{cc} 0 & 1 \\ 0 & 0 \\ \end{array} \right), \hspace{12pt}
  \hat{\sigma}_- = \left( \begin{array}{cc} 0 & 0 \\ 1 & 0 \\ \end{array} \right),
\end{equation}
and \( \hat{\sigma}^{(i)}_{x,z,+,-} = \mathbbm{1}^{[1]} \otimes \mathbbm{1}^{[2]} \dots \otimes \hat{\sigma}^{[i]}_{x,z,+,-} \dots \otimes \mathbbm{1}^{[N]} \).
This transformation transforms our original Hamiltonian, i.e.
\begin{equation}
\hat{H} =
\sum_{i=1}^{N} - \frac{\Delta_i}{2} \hat{\sigma}_z^{(i)} +
\sum_{i=1}^{N-1} J_{i,i+1} \hat{\sigma}_x^{(i)} \hat{\sigma}_x^{(i+1)}
,
\label{eq:chain}
\end{equation}
into
\begin{equation}
\hat{H} =
\sum_{i=1}^{N} \Delta_i \hat{c}_i^{\dagger} \hat{c}_i +
\sum_{i=1}^{N-1} J_{i,i+1} \left( \hat{c}_i^{\dagger} \hat{c}_{i+1} +
\hat{c}_{i+1}^{\dagger} \hat{c}_i + \hat{c}_i^{\dagger}
\hat{c}_{i+1}^{\dagger} + \hat{c}_{i+1} \hat{c}_i \right)
+ {\rm const.}
\label{eq:HamScalc}
\end{equation}
This Hamiltonian does not look much simpler than the original one.
In order to obtain something simple out of it, we now write the Hamiltonian as
\begin{equation}
\hat{H} = \left( \hat{c}_1^{\dagger}, \cdots ,
\hat{c}_N^{\dagger}, \hat{c}_1 \cdots, \hat{c}_N \right) M \left(
\hat{c}_1, \cdots , \hat{c}_N, \hat{c}_1^{\dagger}, \cdots ,
\hat{c}_N^{\dagger} \right)^{\rm transpose} + {\rm const.},
\end{equation}
where $M$ is a $2N \times 2N$ matrix. It is always possible to
define new operators $\hat{\gamma}_i$ (with $i=1,2,...,N$) that
also obey Fermi statistics and for which the Hamiltonian takes the
simple form
\begin{equation}
\hat{H} = \left( \hat{\gamma}_1^{\dagger}, \cdots ,
\hat{\gamma}_N^{\dagger}, \hat{\gamma}_1, \cdots , \hat{\gamma}_N
\right) \tilde{M} \left( \hat{\gamma}_1, \cdots , \hat{\gamma}_N,
\hat{\gamma}_1^{\dagger}, \cdots , \hat{\gamma}_N^{\dagger}
\right)^{\rm transpose} + {\rm const.},
\end{equation}
with a diagonal matrix $\tilde{M}$ (note that this step can be
carried out by diagonalizing the matrix $M$). In other words,
\begin{equation}
\hat{H} = \sum_{i=1}^{N} E_i \hat{\gamma}_i^{\dagger}
\hat{\gamma}_i + {\rm const.}
\end{equation}

The physical meaning of the above Hamiltonian is the following.
We have defined new creation and annihilation operators, which are
combinations of the creation and annihilation operators
for the original fermions. With the new definition, the
Hamiltonian has turned into a Hamiltonian of non-interacting
fermions. By diagonalizing the matrix $M$, we therefore obtain the
available single-particle energy levels for these newly defined
particles. This transformation can be interpreted as going from a
description using the bare qubit states to a picture where the states
are dressed by the qubit-qubit coupling energies.
For a system with fixed qubit-qubit couplings, the dressed states
are the most natural and physically relevant states to describe the
dynamics~\cite{galiautdinov:11}.

The total energy of the entire system is obtained by
identifying which single-particle energy levels are occupied and
taking the sum of the energies of these levels. In other words, any energy
eigenstate can be expressed as $\hat{\gamma}_i^{\dagger}
\hat{\gamma}_j^{\dagger} \cdots \hat{\gamma}_k^{\dagger} \ket{G}$,
where $i,j,...,k$ denote the occupied single-particle states and
$\ket{G}$ is the ground state of the system.
This derivation
provides a somewhat intuitive explanation to the result
of reference \cite{ashhab:08},
that the Larmor frequency of a given qubit is independent of the
states of the other qubits. Since the original Hamiltonian is
equivalent to a Hamiltonian that describes non-interacting
fermions, it is natural that the change in energy induced by the
addition of one particle into a certain energy level is
independent of whether the other energy levels are occupied or
not. This statement is independent of the strength of the
coupling: even for strong coupling, where the energy eigenstates of the
spin system are very involved (which corresponds
to the fermion energy eigenstates having a large extension in space),
the symmetric energy-level structure is still preserved. The
only requirement is a one-dimensional chain with a Hamiltonian of
the form given in equation~(\ref{eq:chain}).

Now let us turn to the eigenstates of the Hamiltonian. In general,
determining these eigenstates is harder than determining the
eigenvalues, because even the ground state has a complicated form.
The ground state $\ket{G}$ is defined by the set of conditions
\begin{equation}
\hat{\gamma}_i \ket{G} = 0
\end{equation}
for all values of $i$. In other words, there are no $\hat{\gamma}$
fermions in the ground state. However, in the language of the
$\hat{c}$ fermions, the above set of equations are not transparent
at all and do not have any simple solution in general. This
difficulty could make calculations complicated. However, we
are mostly interested in the practical situation where the
typical scale of the coupling energies ($J$) is
small compared to the typical energy scale of the difference between different
qubit splittings ($\delta\Delta$). We can therefore perform
perturbation-theory calculations where we start by considering the
case $J=0$ first and then analyse the effect of adding a small
coupling term.

In the case of zero coupling ($J=0$), the matrix $M$ is already
diagonal, which means that the $\hat{\gamma}$ fermions coincide with
the $\hat{c}$ fermions. This means that each
(single-particle) energy level is localized at one lattice site
with no tunneling between these sites. The state of the entire
system is defined simply by specifying the occupation (0 or 1) of
the different lattice sites. We now add a small coupling term
($J\neq 0$). For simplicity, we first consider the
condition that the coupling strength is very small compared to the
qubit Larmor frequencies ($J\ll\Delta$; note that here we use
$\Delta$, not $\delta\Delta$) and make the rotating wave
approximation, meaning that the terms in equation~(\ref{eq:HamScalc})
that create or annihilate two $\hat{c}$ fermions are ignored:
\begin{equation}
\hat{H} \approx
\sum_{i=1}^{N} \Delta_i \hat{c}_i^{\dagger} \hat{c}_i +
\sum_{i=1}^{N-1} J_{i,i+1} \left(
\hat{c}_i^{\dagger} \hat{c}_{i+1} + \hat{c}_{i+1}^{\dagger}
\hat{c}_i \right) + {\rm const.}
\end{equation}
This approximation leads to a conceptual simplification, because
we now know that creating a $\hat{\gamma}$ fermion in a given
energy level corresponds to creating a $\hat{c}$ fermion in a
quantum state that is (in general) spread over the entire lattice,
with no involved mixing between creation and annihilation operators.
Another important result of the above approximation is that the
ground state reduces to a very simple form. Regardless of the
exact form of the operators $\hat{\gamma}_i$ in terms of the
operators $\hat{c}_i$, the set of conditions $\hat{\gamma}_i
\ket{G}=0$ leads to the set of conditions $\hat{c}_i \ket{G}=0$,
such that the ground state corresponds to zero occupation of all
the single-particle states in both languages. Note that the number
of particles is also the same for any state in both languages
(\(\sum \hat{c}^\dagger \hat{c} = \sum \hat{\gamma}^\dagger \hat{\gamma}\)).

With the above approximation, we have turned the Hamiltonian into
a Hamiltonian that conserves the number of particles (even in the
language of the $\hat{c}$ operators). Note also that there are no
interaction terms in the Hamiltonian. We can therefore proceed in
calculating the form of the operators $\hat{\gamma}_i$ by
considering a single-particle problem (This is allowed because in
the absence of interactions the form of the different
single-particle wave functions will be insensitive to whether
other wave functions are occupied or not). For this purpose we
imagine that there is only one particle in the system, and this
particle can hop (i.e. tunnel) between the different lattice sites.

When calculating the form of the single-particle wave functions,
we use the condition $J\ll\delta\Delta$. This condition implies
that although the $\hat{c}$ fermions can hop between the different
lattice sites, the hopping matrix elements are small compared to
the detuning between neighbouring sites. As a result, the
different single-particle states (i.e., those described by the
$\hat{\gamma}$ fermions) will each be concentrated around one
lattice site (recall that when $J=0$ the states are completely
localized at individual lattice sites). One can calculate the
spread of the states using perturbation theory. Writing
\begin{equation}
\hat{\gamma}_i = \sum_{j=1}^N \psi_i(j) \hat{c}_j,
\end{equation}
we find that:
\begin{eqnarray}
\psi_i(i) & = & 1 + O \left( \frac{J}{\delta\Delta} \right)^2 \nonumber \\
\psi_i(i \pm 1) & = & \frac{J_{i,i\pm 1}}{\Delta_i-\Delta_{i\pm
1}} + O \left( \frac{J}{\delta\Delta} \right)^3 \nonumber \\
\psi_i(i \pm 2) & = & \frac{J_{i,i\pm 1} J_{i\pm 1,i\pm
2}}{(\Delta_i-\Delta_{i\pm 1})(\Delta_i-\Delta_{i\pm 2})} + O
\left( \frac{J}{\delta\Delta} \right)^4 \nonumber \\
& \vdots & \nonumber \\
\psi_i(j) & \sim & \left( \frac{J}{\delta\Delta} \right)^{|i-j|}
\hspace{1cm} {\rm when \ } |i-j| \gg 1 \nonumber \\
& \vdots &
\end{eqnarray}
Note that the above relation can be inverted straightforwardly:
\begin{equation}
\hat{c}_i = \sum_{j=1}^N \psi_j(i) \hat{\gamma}_j.
\end{equation}
In principle the localization situation changes when two or more
lattice sites are degenerate. For example, in the case of two-fold
degeneracy and an otherwise symmetric environment,
the (single-particle) energy eigenstates will
be symmetric and antisymmetric superpositions of the
particle being at one of the two sites. However, provided that
these degeneracies occur only for largely separated lattice sites,
the effective tunneling matrix element will be exponentially small
in the distance between the two sites in question: (assuming
$j>i$)
\begin{equation}
J_{ij,\rm effective} \sim \frac{J_{i,i+1}}{\Delta_i-\Delta_{i+1}}
\times \frac{J_{i+1,i+2}}{\Delta_i-\Delta_{i+2}} \times \cdots
\frac{J_{j-2,j-1}}{\Delta_i-\Delta_{j-1}} \times J_{j-1,j}.
\end{equation}
We can therefore ignore such long-range tunneling and focus on the
practically relevant case where the $\hat{\gamma}$ fermions and
the $\hat{c}$ fermions are almost equal, with exponentially
decaying tails describing the spread of the $\hat{\gamma}$
fermions. This situation is related to the phenomenon of Anderson
localization, where disorder causes the single-particle states to
be localized with exponentially decaying tails.

We can now start doing the calculations for the Rabi frequencies.
Before going into long expressions, we make the following
observation. If a driving signal is applied to qubit $i$, the
driving term in the Hamiltonian can be expressed in the language
of fermions as:
\begin{equation}
\hat{H}_{\rm drive} =
\prod_{j<i} \left( 1 - 2
\hat{c}_j^{\dagger} \hat{c}_j \right) \left( \hat{c}_i +
\hat{c}_i^{\dagger} \right) \cos(\omega t).
\end{equation}
The relevant matrix elements for purposes of evaluating the Rabi
frequency of qubit \( k \) therefore have the form:
\begin{equation}
\bra{0} \hat{\gamma}_m \hat{\gamma}_n \cdots \hat{\gamma}_p \left[
\hat{\gamma}_k \prod_{j<i} \left( 1 - 2 \hat{c}_j^{\dagger}
\hat{c}_j \right) \left( \hat{c}_i + \hat{c}_i^{\dagger} \right)
\right] \hat{\gamma}_p^{\dagger} \cdots \hat{\gamma}_n^{\dagger}
\hat{\gamma}_m^{\dagger} \ket{0},
\end{equation}
where the index $i$ represents the driven qubit and the operators
on the far right and far left (i.e. those with indices
$m,n,...,p$) describe the state of the surrounding qubits. Let us
now assume that $i$ and $m$ are separated by a large distance.
From the localization argument above, we know that the fermion
created by the operator $\hat{\gamma}_m^{\dagger}$ is concentrated
mostly on one side of site $i$ (i.e. to the right if $m>i$ and to
the left if $m<i$), with only a very small probability of
occupying site $i$ or crossing to the opposite side. As mentioned
above, this probability decreases exponentially with the distance
between $i$ and $m$. We can therefore make the approximation that
the operators $\left( \hat{c}_i + \hat{c}_i^{\dagger} \right)$ and
$\hat{\gamma}_m^{\dagger}$ anti-commute and that the operators
$\prod_{j<i} \left( 1 - 2 \hat{c}_j^{\dagger} \hat{c}_j \right)$
and $\hat{\gamma}_m^{\dagger}$ either commute or anti-commute,
depending on whether $m>i$ or $m<i$. Note here that the operator
$\prod_{j<i} \left( 1 - 2 \hat{c}_j^{\dagger} \hat{c}_j \right)$
counts the number of particles to the left of site $i$ and
produces a sign factor $\pm 1$ depending on that number. This
approximation, along with the anti-commutation relations between
the different $\hat{\gamma}$ operators, allow us to move the
operator $\hat{\gamma}_m^{\dagger}$ from the far right of the
above expression to just right of the operator $\hat{\gamma}_m$,
with only a minus sign appearing whenever $m<i$. The combination
$\hat{\gamma}_m \hat{\gamma}_m^{\dagger}$ applied to $\bra{0}$
does not affect the state, and these two operators can be ignored.
We therefore conclude that the effect of the occupation of a
distant single-particle state is exponentially small and can be
ignored. We now have to decide where to draw the line between
states that we will consider and states that we will ignore. Since
we are interested in lowest-order corrections, we shall consider
only the effect of nearest-neighbour states, which will
produce the lowest-order corrections.

As a first step, let us consider the Rabi frequency of qubit $i$
when driving only that qubit. Since under our approximation we
only consider the effect of the occupation of the two neighboring
sites, we only need to consider a three-qubit system (the algebra
can be easily generalized to longer chains).
It is straightforward to show that for driving qubit 1 we
have
\begin{eqnarray}
\bra{0} \hat{\gamma}_1 \left(\hat{c}_1+\hat{c}_1^{\dagger}\right)
\ket{0} & = & \bra{0} \hat{\gamma}_2 \hat{\gamma}_1
\left(\hat{c}_1+\hat{c}_1^{\dagger}\right)
\hat{\gamma}_2^{\dagger} \ket{0} = \bra{0} \hat{\gamma}_3
\hat{\gamma}_1 \left(\hat{c}_1+\hat{c}_1^{\dagger}\right)
\hat{\gamma}_3^{\dagger} \ket{0} = \bra{0} \hat{\gamma}_2
\hat{\gamma}_3 \hat{\gamma}_1
\left(\hat{c}_1+\hat{c}_1^{\dagger}\right)
\hat{\gamma}_3^{\dagger}
\hat{\gamma}_2^{\dagger} \ket{0} \nonumber \\
& = & \psi_1(1),
\end{eqnarray}
and a similar relation applies to driving qubit 3. For driving
qubit 2 we have
\begin{subequations}
\begin{eqnarray}
\bra{0} \hat{\gamma}_2 \left( 1 - 2 \hat{c}_1^{\dagger} \hat{c}_1
\right) \left( \hat{c}_2 + \hat{c}_2^{\dagger} \right) \ket{0}
& = & \psi_2(2)
\\
- \bra{0} \hat{\gamma}_1 \hat{\gamma}_2 \left( 1 - 2
\hat{c}_1^{\dagger} \hat{c}_1 \right)
\left(\hat{c}_2+\hat{c}_2^{\dagger}\right)
\hat{\gamma}_1^{\dagger} \ket{0}
& \approx & \psi_2(2) \left[ 1 + \frac{2 J_{12}^2 J_{23}^2}
{(\Delta_1-\Delta_2)(\Delta_2-\Delta_3)(\Delta_1-\Delta_3)^2}
\right]
\\
\bra{0} \hat{\gamma}_3 \hat{\gamma}_2 \left( 1 - 2
\hat{c}_1^{\dagger} \hat{c}_1 \right)
\left(\hat{c}_2+\hat{c}_2^{\dagger}\right)
\hat{\gamma}_3^{\dagger} \ket{0}
& \approx & \psi_2(2) \left[ 1 + \frac{2 J_{12}^2 J_{23}^2}
{(\Delta_1-\Delta_2)(\Delta_2-\Delta_3)(\Delta_1-\Delta_3)^2}
\right]
\\
- \bra{0} \hat{\gamma}_1 \hat{\gamma}_3 \hat{\gamma}_2 \left( 1 -
2 \hat{c}_1^{\dagger} \hat{c}_1 \right) \left( \hat{c}_2 +
\hat{c}_2^{\dagger} \right) \hat{\gamma}_3^{\dagger}
\hat{\gamma}_1^{\dagger} \ket{0}
& = & \psi_2(2).
\end{eqnarray}
\end{subequations}
Therefore the Rabi frequencies of qubits 1 and 3 (which are at the
end of the chain) do not depend on the state of the other qubits
(this result holds even for longer chains and can be shown
rigorously using Wick's theorem), while the Rabi frequency of
qubit 2 shows some dependence on the states of the surrounding
qubits, with a spread that is of the order of
$(J/\delta\Delta)^4$. These results agree with the numerically
derived results of reference \cite{ashhab:08}.

We now turn to the Rabi frequencies when driving two qubits
simultaneously in order to implement the SD-CNOT gate.
We now have
driving terms applied to two qubits. As a result, in addition to
the expressions calculated above, we need to evaluate expressions
of the form
\begin{equation}
\bra{0} \hat{\gamma}_m \hat{\gamma}_n \cdots \hat{\gamma}_p \left[
\hat{\gamma}_i \prod_{j\leq i} \left( 1 - 2 \hat{c}_j^{\dagger}
\hat{c}_j \right) \left( \hat{c}_{i+1} + \hat{c}_{i+1}^{\dagger}
\right) \right] \hat{\gamma}_p^{\dagger} \cdots
\hat{\gamma}_n^{\dagger} \hat{\gamma}_m^{\dagger} \ket{0}.
\label{eq:scaldriveother}
\end{equation}
As above, we ignore all the distant qubits. We consider a
four-qubit chain with the driving applied to qubit 3 such that
qubit 2 changes its state [i.e. $i=2$ in equation~(\ref{eq:scaldriveother})]. Here we take a chain of length four, to also incorporate the influence of the
nearest-neighbour of the driven qubit. The relevant
matrix elements are given by
\begin{subequations}
\begin{eqnarray}
\bra{0} \hat{\gamma}_2 \prod_{j\leq 2} \left( 1 - 2
\hat{c}_j^{\dagger} \hat{c}_j \right) \left( \hat{c}_3 +
\hat{c}_3^{\dagger} \right) \ket{0}
& = & \psi_2(3)
\\
- \bra{0} \hat{\gamma}_1 \hat{\gamma}_2 \prod_{j\leq 2} \left( 1 -
2 \hat{c}_j^{\dagger} \hat{c}_j \right) \left( \hat{c}_3 +
\hat{c}_3^{\dagger} \right) \hat{\gamma}_1^{\dagger} \ket{0}
& = & \psi_2(3) + 2 \psi_1(4) \times \nonumber
\\
& & \hspace{0.0cm} \frac{J_{12} J_{23}^2
J_{34}}{(\Delta_1-\Delta_3)(\Delta_2-\Delta_3)(\Delta_2-\Delta_4)
(\Delta_1-\Delta_4)}
\\
\bra{0} \hat{\gamma}_4 \hat{\gamma}_2 \prod_{j\leq 2} \left( 1 - 2
\hat{c}_j^{\dagger} \hat{c}_j \right) \left( \hat{c}_3 +
\hat{c}_3^{\dagger} \right) \hat{\gamma}_4^{\dagger} \ket{0}
& \approx & \psi_2(3) + 2 \psi_4(2) \psi_4(3) \psi_2(2)
\\
- \bra{0} \hat{\gamma}_1 \hat{\gamma}_4 \hat{\gamma}_2
\prod_{j\leq 2} \left( 1 - 2 \hat{c}_j^{\dagger} \hat{c}_j \right)
\left( \hat{c}_3 + \hat{c}_3^{\dagger} \right)
\hat{\gamma}_4^{\dagger} \hat{\gamma}_1^{\dagger} \ket{0}
& \approx & \psi_2(3) + 2 \psi_4(2) \psi_4(3) \psi_2(2),
\end{eqnarray}
\end{subequations}
and similar expressions for the case where the operator
$\hat{\gamma}_3$ appears, e.g.
\begin{eqnarray}
- \bra{0} \hat{\gamma}_1 \hat{\gamma}_3 \hat{\gamma}_4
\hat{\gamma}_2 \prod_{j\leq 2} \left( 1 - 2 \hat{c}_j^{\dagger}
\hat{c}_j \right) \left( \hat{c}_3 + \hat{c}_3^{\dagger} \right)
\hat{\gamma}_4^{\dagger} \hat{\gamma}_3^{\dagger}
\hat{\gamma}_1^{\dagger} \ket{0}
& = & \psi_2(3),
\end{eqnarray}

We therefore arrive at the main result of this section
\begin{subequations}
\begin{eqnarray}
\bra{0} \hat{\gamma}_2 \prod_{j\leq 2} \left( 1 - 2
\hat{c}_j^{\dagger} \hat{c}_j \right) \left( \hat{c}_3 +
\hat{c}_3^{\dagger} \right) \ket{0} & = & \psi_2(3) \\
- \bra{0} \hat{\gamma}_1 \hat{\gamma}_2 \prod_{j\leq 2} \left( 1 -
2 \hat{c}_j^{\dagger} \hat{c}_j \right) \left( \hat{c}_3 +
\hat{c}_3^{\dagger} \right) \hat{\gamma}_1^{\dagger} \ket{0} & = &
\psi_2(3) \left( 1 + O \left( \frac{J}{\delta\Delta} \right)^6
\right) \\
\bra{0} \hat{\gamma}_4 \hat{\gamma}_2 \prod_{j\leq 2} \left( 1 - 2
\hat{c}_j^{\dagger} \hat{c}_j \right) \left( \hat{c}_3 +
\hat{c}_3^{\dagger} \right) \hat{\gamma}_4^{\dagger} \ket{0} & = &
\psi_2(3) \left( 1 + O \left( \frac{J}{\delta\Delta} \right)^2
\right) \\
- \bra{0} \hat{\gamma}_1 \hat{\gamma}_4 \hat{\gamma}_2
\prod_{j\leq 2} \left( 1 - 2 \hat{c}_j^{\dagger} \hat{c}_j \right)
\left( \hat{c}_3 + \hat{c}_3^{\dagger} \right)
\hat{\gamma}_4^{\dagger} \hat{\gamma}_1^{\dagger} \ket{0} & = &
\psi_2(3) \left( 1 + O \left( \frac{J}{\delta\Delta} \right)^2
\right)
\end{eqnarray}
\end{subequations}
and similar expressions for the case where the operator
$\hat{\gamma}_3$ appears.

The relative error in the CNOT gate will therefore be of order
$(J/\delta\Delta)^2$. This result is in agreement with the
numerical calculations in reference~\cite{ashhab:08} where also
a spread of $(J/\delta\Delta)^2$ is found in the matrix elements
for flipping qubit $i$ when driving
qubit $i+1$, depending on the states of the surrounding qubits.

The scaling of the error with $(J/\delta\Delta)^2$ is comparable to other
schemes for implementing entangling two-qubit gates.
The SD-CNOT gate has however the added advantage that the qubit level
splittings do not need to be shifted. The shifting of energy levels gets
increasingly complicated in multi-qubit systems, where the multitude of
levels makes it harder to avoid crossing resonances that lead to unwanted
evolution and entanglement. Also the freedom that the qubits do not need
to be nearest-neighbour in frequency can make the design of a multi-qubit
system simpler.

\section{Conclusion and discussion}
We have theoretically analysed all the properties of the SD-CNOT gate that are important for its implementation in experiments. In the weak-driving limit, and in the absence of decoherence, the fidelity of the gate is 100\%. For increasing driving strength the fidelity of the gate decreases due to the off-resonant driving of other transitions in the system. The off-resonant driving can induce spurious excitation of these transitions and AC-stark shifts of the levels, inducing an undesired single-qubit and (entangling) two-qubit evolution. We analysed the strength of these effects analytically, and in addition performed a full numerical simulation of the time evolution of the system. For the SD-CNOT gate without extra optimizations, the numerical simulations show a fidelity of 99\% and 99.99\% for gate speeds corresponding to \(0.48J\) and \(0.07J\) respectively. At high gate speeds the fidelity can be boosted by employing single-qubit correction pulses before and after the two-qubit driving.

  The gate performance was also analysed for multi-level qubits with weak anharmonicity of the energy splittings.
  As expected, it was found that the gate speed for a specified fidelity goes down for decreasing anharmonicity of the qubits. The energy shifts of the computational states due to the higher levels and leakage to these higher levels can both play an important role. We determined fidelities of 99\% and 99.99\% for gate speeds corresponding to \( 0.08\delta \) and \( 0.02\delta \) (for \( \delta\approx200~\mathrm{MHz}\)). The gate speed can be increased by designing a relatively large coupling strength or low detuning between the qubits, or both. Another possible route towards increasing the gate speed and/or fidelity for weakly anharmonic qubits is using pulse shaping of the driving field.

  The scalability of the SD-CNOT gate was also studied. For a 1-dimensional array of qubits the error in the gate scales as \( (J/\delta\Delta)^2 \), which is comparable to other schemes for implementing entangling two-qubit gates.
The SD-CNOT gate has however the favorable properties that the qubit level
splittings do not need to be shifted, and neither is it required that
the qubits are nearest-neighbour in frequency. These properties can make
a significant difference in the design of multi-qubit systems.
  We conclude that in terms of fidelity, gate speed and potential for scalability, the SD-CNOT gate provides a valuable addition to the existing variety of entangling gates.

\begin{acknowledgments}
 We thank D. Rist\`{e} for discussions.
 PCdG, CJPMH and JEM were supported by NanoNed and FOM and EU projects EuroSQIP and CORNER.
 SA and FN were partially supported by ARO, NSF grant No. 0726909, JSPS-RFBR contract No. 12-02-92100, Grant-in-Aid for Scientific Research (S), MEXT Kakenhi on Quantum Cybernetics, and the JSPS via its FIRST program.
 AL was supported by NSERC.
 LDC was partially funded by NWO.
\end{acknowledgments}

\appendix

\section{Transformation to rotating frame}\label{apx:rotframe}
We start from the Hamiltonian and driving term as given by equations~(\ref{eq:Hamiltonian}) and (\ref{eq:Hdrive}). For this calculation we use the energy eigenbasis in the absence
of the driving term [i.e.~the states $\ket{0},\ket{1},\ket{2}$ and
$\ket{3}$ in equations~(\ref{eq:eigv2})]. The Hamiltonian without the
driving term takes the form:
\begin{equation}
  \hat{H}_0 =
\frac{1}{2} \left( \begin{array}{cccc}
-\sqrt{(\Delta_1+\Delta_2)^2+4J^2} & 0
& 0 & 0 \\
0 & - \sqrt{(\Delta_1-\Delta_2)^2+4J^2} & 0
& 0 \\
0 & 0 & \sqrt{(\Delta_1-\Delta_2)^2+4J^2} & 0 \\
0 & 0 & 0 & \sqrt{(\Delta_1+\Delta_2)^2+4J^2} \\
\end{array}
\right)
\label{eq:Htimedom}
\end{equation}
and the matrix elements of the driving term  $\hat{H}_{\rm drive} $ [equation~(\ref{eq:Hdrive})] are given by
\begin{eqnarray}
\bra{1} \hat{H}_{\rm drive} \ket{0} & = & -a_1
\sin\theta_+ \cos(\omega t + \varphi_1) + a_2
\cos\theta_- \cos(\omega t + \varphi_2) \nonumber
\\
\bra{2} \hat{H}_{\rm drive} \ket{0} & = & + a_1
\cos\theta_+ \cos(\omega t + \varphi_1) + a_2
\sin\theta_- \cos(\omega t + \varphi_2) \nonumber
\\
\bra{3} \hat{H}_{\rm drive} \ket{2} & = & + a_1
\sin\theta_+ \cos(\omega t + \varphi_1) + a_2
\cos\theta_- \cos(\omega t + \varphi_2) \nonumber
\\
\bra{3} \hat{H}_{\rm drive} \ket{1} & = & + a_1
\cos\theta_+ \cos(\omega t + \varphi_1) - a_2
\sin\theta_- \cos(\omega t + \varphi_2) \nonumber
\\
\bra{i} \hat{H}_{\rm drive} \ket{j} & = & \bra{j} \hat{H}_{\rm drive} \ket{i}.
\end{eqnarray}
All other matrix elements vanish.

We now make a transformation to a convenient rotating frame. This
means that instead of solving the Schr\"odinger equation for
$\ket{\psi}$, we solve an equation for
\begin{equation}
\ket{\widetilde{\psi}} = \hat{U}(t) \ket{\psi},
\label{eq:phitilde}
\end{equation}
where
\begin{equation}
  \hat{U}(t) = e^{iut / \hbar}, \hspace{12pt}
u = \left( \begin{array}{cccc}
- \hbar \omega & 0 & 0 & 0 \\
0 & 0 & 0 & 0 \\
0 & 0 & 0 & 0 \\
0 & 0 & 0 & \hbar \omega \\
\end{array}
\right).
\end{equation}
Note that \( u \) is not the only possible choice that could serve our purpose here. Other transformations will lead to the same result.

The lab-frame state $\ket{\psi}$ satisfies the Schr\"odinger
equation: \( i \hbar \frac{d}{dt} \ket{\psi} = \hat{H} \ket{\psi} \). In the rotating frame,
the state $\ket{\widetilde{\psi}}$ should also satisfy the Schr\"odinger equation:
\( i \hbar \frac{d}{dt} \ket{\widetilde{\psi}} = \tilde{H} \ket{\widetilde{\psi}}\). From these equations the transformation is easily derived:
\begin{eqnarray}
\tilde{H} & = & \hat{U} \hat{H} \hat{U}^{\dagger} - i \hbar \hat{U}
\frac{d\hat{U}^{\dagger}}{dt} \nonumber
\\
& = & \hat{U} \hat{H} \hat{U}^{\dagger} - u.
\end{eqnarray}
With the total Hamiltonian given by \( \hat{H} = \hat{H}_0 + \hat{H}_{\rm drive} \), this results in
\begin{subequations}
\begin{eqnarray}
\tilde{H}_0 & = & \hat{U} \hat{H}_0 \hat{U}^{\dagger} - u
\nonumber
\\
& = & \hat{H}_0 - u \nonumber
\\
& = & \left( \begin{array}{cccc} -
\frac{\sqrt{(\Delta_1+\Delta_2)^2+4J^2}}{2} + \hbar\omega & 0
& 0 & 0 \\
0 & -  \frac{\sqrt{(\Delta_1-\Delta_2)^2+4J^2}}{2} & 0
& 0 \\
0 & 0 &  \frac{\sqrt{(\Delta_1-\Delta_2)^2+4J^2}}{2} & 0 \\
0 & 0 & 0 &  \frac{\sqrt{(\Delta_1+\Delta_2)^2+4J^2}}{2} - \hbar\omega \\
\end{array}
\right)
\\
\tilde{H}_{\rm drive} & = & \hat{U} \hat{H}_{\rm drive} \hat{U}^{\dagger}.
\end{eqnarray}
\end{subequations}
The matrix elements of $\tilde{H}_{\rm drive}$ can be conveniently split into $\tilde{H}_{\rm drive} = \tilde{V} + \tilde{W}$, where $\tilde{V}$ contains the time-independent and $\tilde{W}$ the time-dependent terms:
\begin{eqnarray}
\bra{1} \tilde{V} \ket{0} & = & e^{i\omega t} \bra{1} \hat{V}
\ket{0} \nonumber
\\
& = & \frac{1}{2} \left(
- a_1 \sin\theta_+ e^{-i\varphi_1}
+ a_2 \cos\theta_- e^{-i\varphi_2} \right)
\nonumber
\\
\bra{2} \tilde{V} \ket{0} & = & \frac{1}{2} \left(
+ a_1 \cos\theta_+ e^{-i\varphi_1}
+ a_2 \sin\theta_- e^{-i\varphi_2} \right)
\nonumber
\\
\bra{3} \tilde{V} \ket{2} & = & \frac{1}{2} \left(
+ a_1 \sin\theta_+ e^{-i\varphi_1}
+ a_2 \cos\theta_- e^{-i\varphi_2} \right)
\nonumber
\\
\bra{3} \tilde{V} \ket{1} & = & \frac{1}{2} \left(
+ a_1 \cos\theta_+ e^{-i\varphi_1}
- a_2 \sin\theta_- e^{-i\varphi_2} \right)
\nonumber
\\
\bra{i} \tilde{V} \ket{j} & = & \bra{j} \tilde{V} \ket{i}^*
\nonumber
\\
\bra{1} \tilde{W} \ket{0} & = & \frac{1}{2} \left(
- a_1 \sin\theta_+ e^{i\varphi_1}
+ a_2 \cos\theta_- e^{i\varphi_2} \right)
e^{i2\omega t} \nonumber
\\
\bra{2} \tilde{W} \ket{0} & = & \frac{1}{2} \left(
+ a_1 \cos\theta_+ e^{i\varphi_1}
+ a_2 \sin\theta_- e^{i\varphi_2} \right)
e^{i2\omega t} \nonumber
\\
\bra{3} \tilde{W} \ket{2} & = & \frac{1}{2} \left(
+ a_1 \sin\theta_+ e^{i\varphi_1}
+ a_2 \cos\theta_- e^{i\varphi_2} \right)
e^{i2\omega t} \nonumber
\\
\bra{3} \tilde{W} \ket{1} & = & \frac{1}{2} \left(
+ a_1 \cos\theta_+ e^{i\varphi_1}
- a_2 \sin\theta_- e^{i\varphi_2} \right)
e^{i2\omega t} \nonumber
\\
\bra{i} \tilde{W} \ket{j} & = & \bra{j} \tilde{W} \ket{i}^*.
\label{eq:VWtilde}
\end{eqnarray}
Transforming back to the lab-frame, the matrix elements in the energy eigenbasis of $\hat{V}$ and $\hat{W}$ are given by
\begin{eqnarray}
\bra{1} \hat{V} \ket{0} & = & \frac{1}{2} \left(
- a_1 \sin\theta_+ e^{-i\varphi_1}
+ a_2 \cos\theta_- e^{-i\varphi_2} \right)
e^{-i\omega t} \nonumber
\\
\bra{2} \hat{V} \ket{0} & = & \frac{1}{2} \left(
+ a_1 \cos\theta_+ e^{-i\varphi_1}
+ a_2 \sin\theta_- e^{-i\varphi_2} \right)
e^{-i\omega t} \nonumber
\\
\bra{3} \hat{V} \ket{2} & = & \frac{1}{2} \left(
+ a_1 \sin\theta_+ e^{-i\varphi_1}
+ a_2 \cos\theta_- e^{-i\varphi_2} \right)
e^{-i\omega t} \nonumber
\\
\bra{3} \hat{V} \ket{1} & = & \frac{1}{2} \left(
+ a_1 \cos\theta_+ e^{-i\varphi_1}
- a_2 \sin\theta_- e^{-i\varphi_2} \right)
e^{-i\omega t} \nonumber
\\
\bra{i} \hat{V} \ket{j} & = & \bra{j} \hat{V} \ket{i}^* \nonumber
\\
\bra{i} \hat{W} \ket{j} & = & \bra{i} \hat{V} \ket{j}^*.
\label{eq:VWeb}
\end{eqnarray}
From these equations we see that $\hat{V}$ describes the co-rotating and $\hat{W}$ the counter-rotating field with respect to the Larmor precession of the system.

The representation of $\hat{V}$ in the uncoupled-qubit basis can be straightforwardly calculated from the result of equation~(\ref{eq:VWeb}), however the result is a rather involved and large matrix. The main results of this appendix, i.e. the matrix elements in equation~(\ref{eq:VWtilde}),  can be reproduced exactly by use of the terms $\tilde{H}_{\rm drive}^+$ and $\tilde{H}_{\rm drive}^-$ as described in subsection~\ref{sec:seldaA}. This approach is valid for the purpose of calculating transition strengths, but for other purposes it can be necessary to use the full representation $\hat{V}$ and $\hat{W}$.

Note that in the derivation in this appendix no approximations have been
made: we have only performed transformations and rearranged equations.

\section{Quantum mechanical description of the driving field}\label{apx:maxspeed}

In the other sections we have treated the driving field as a classical field.
In this section we provide a description of the driven system where the driving
field is treated quantum mechanically. With this description we verify some of the
results on level shifts and gate errors obtained using the classical field description.
Additionally, the description provided here can be used as a starting point
to further investigate the strong-driving regime, maximum gate speed and the
related gate errors.

For this calculation we use the
dressed-state picture. We consider a model composed of
two qubits and a harmonic oscillator containing a large number of
photons. In this picture the Hamiltonian becomes time
independent and the frequency of the driving field now appears as
\( \hbar \omega \hat{b}^{\dagger} \hat{b} \), where $\hat{b}$ and
$\hat{b}^{\dagger}$ are the annihilation and creation operators of
the quantized field. The driving term
\( a_i \cos(\omega t)\) in the original Hamiltonian
(i.e.~the Hamiltonian in which the driving field is treated as
classical) is now replaced by the operator ${g_i}
(\hat{b}+\hat{b}^{\dagger})$, where  \( g_i \) represents the coupling
strength between the resonator and qubit \( i \).

We take the
case where qubit 2 is the target qubit and where the
$\ket{00}\leftrightarrow\ket{01}$ transition is to be darkened. In
this case $\varphi_1=\varphi_2$. Again, for simplicity in the expressions,
we take \( \varphi_1 = \varphi_2 = 0 \).
Unlike in the weak-driving limit, we do
not impose the relations $\hbar\omega=E_1-E_0$ and
\( a_2/a_1 = \sin\theta_+/\cos\theta_- \).
Instead, we leave the coupling strengths between the oscillator
and the two qubits as tunable parameters to be decided later.

Assuming that the qubit frequencies $\Delta_1$ and $\Delta_2$ are
much larger than the coupling and driving frequency scales,
the energy-level structure breaks up into blocks of four energy
levels in each block, with negligible effects caused by the
coupling between the different blocks. We therefore consider the
spaces spanned by the states $\ket{0,N+1}$, $\ket{1,N}$,
$\ket{2,N}$ and $\ket{3,N-1}$.
In this space the Hamiltonian reads:
\begin{equation}
\hat{H} = \left( \begin{array}{cccc}
  0 & -C_1 \sin\theta_+ + C_2 \cos\theta_- &
    C_1 \cos\theta_+ + C_2 \sin\theta_- & 0 \\
  -C_1 \sin\theta_+ + C_2 \cos\theta_- &
    (E_1-E_0)-\hbar\omega & 0 &
    C_1 \cos\theta_+ - C_2 \sin\theta_- \\
  C_1 \cos\theta_+ + C_2 \sin\theta_- & 0 &
    (E_2-E_0)-\hbar\omega &
    C_1 \sin\theta_+ + C_2 \cos\theta_- \\
  0 & C_1 \cos\theta_+ - C_2 \sin\theta_- &
  C_1 \sin\theta_+ + C_2 \cos\theta_- &
    (E_1+E_2-2E_0)-2\hbar\omega
\end{array}
\right).
\label{eq:Hamstrong}
\end{equation}
where \( C_i = \sqrt{N}g_i \), with \( N \) the number of photons in the resonator, and we have made the assumption that \( N \) is large, such that \( N \approx N+1 \).

One can now argue that the
condition for a darkened transition is that the two corresponding energy
levels must be degenerate: what corresponds to driven (Rabi)
evolution in the previous picture can be regarded as Larmor precession in
the current picture, and for Larmor precession to be suppressed
the corresponding states have to be degenerate.

If we make the approximation that the driving amplitudes are small
compared to the detuning, the above Hamiltonian can be
approximated as
\begin{equation}
\hat{H}_0 \approx \left( \begin{array}{cccc}
  0 & -C_1 \sin\theta_+ + C_2 \cos\theta_- & 0 & 0 \\
  -C_1 \sin\theta_+ + C_2 \cos\theta_- & (E_1-E_0)-\hbar\omega & 0 & 0 \\
  0 & 0 & (E_2-E_0)-\hbar\omega & C_1 \sin\theta_+ + C_2 \cos\theta_- \\
  0 & 0 & C_1 \sin\theta_+ + C_2 \cos\theta_- & (E_1+E_2-2E_0)-2\hbar\omega
\end{array}
\right),
\end{equation}
and then it is obvious that the lowest two energy levels are
degenerate for the choice of parameters $(E_1-E_0)-\hbar\omega=0$ and
$-C_1 \sin\theta_+ + C_2 \cos\theta_-=0$,
which is just a different way of writing the condition derived in
the semiclassical derivation [equation~(\ref{eq:sd_condition_strongdr})].

When we do not make the approximation of small driving amplitudes,
using the full Hamiltonian in equation~(\ref{eq:Hamstrong}),
the matrix elements $H_{02}$, $H_{20}$,
$H_{13}$ and $H_{31}$ will mix all the
quantum states to form the new, corrected energy eigenstates. If
we think in terms of an order-by-order expansion of the effects of
these matrix elements, we see that the matrix elements $H_{02}$
and $H_{20}$ will cause mixing and  repulsion between the states
$\ket{0,N+1}$ and $\ket{2,N}$, while the matrix elements $H_{13}$
and $H_{31}$ will cause mixing and repulsion between the states
$\ket{1,N}$ and $\ket{3,N-1}$. To first order, the corrected
quantum states are:
\begin{subequations}
\begin{eqnarray}
\ket{0,N+1}' & = & \ket{0,N+1} - \frac{C_1 \cos\theta_+
+ C_2 \sin\theta_-}{E_2-E_0-\hbar\omega} \ket{2,N} \\
\ket{1,N}' & = & \ket{1,N} - \frac{C_1 \cos\theta_+ -
C_2 \sin\theta_-}{E_2-E_0-\hbar\omega} \ket{3,N-1} \\
\ket{2,N}' & = & \ket{2,N} + \frac{C_1 \cos\theta_+ +
C_2 \sin\theta_-}{E_2-E_0-\hbar\omega} \ket{0,N+1} \\
\ket{3,N-1}' & = & \ket{3,N-1} + \frac{C_1 \cos\theta_+
- C_2 \sin\theta_-}{E_2-E_0-\hbar\omega} \ket{1,N},
\end{eqnarray}
\end{subequations}
and, to lowest order, the corrections to the energies (\( \delta E_i = E'_i - E_i \)) are given by
\begin{subequations}
\begin{eqnarray}
\delta E_0 & = & - \delta E_2 =
- \frac{\left(C_1 \cos\theta_+ + C_2 \sin\theta_-\right)^2}{E_2-E_0-\hbar\omega} \\
\delta E_1 & = & - \delta E_3 =
- \frac{\left(C_1 \cos\theta_+ - C_2 \sin\theta_-\right)^2}{E_2-E_0-\hbar\omega}.
\end{eqnarray}
\end{subequations}

With the above corrections, we have
\begin{subequations}
\begin{eqnarray}
E'_1-E'_0 & = & E_1-E_0 + \frac{4 C_1 C_2 \cos\theta_+\sin\theta_-}{E_2-E_0-\hbar\omega} \label{eq:E1minE0} \\
E'_3-E'_2 & = & E_3-E_2 - \frac{4 C_1 C_2 \cos\theta_+\sin\theta_-}{E_2-E_0-\hbar\omega} \label{eq:E3minE2} \\
E'_2-E'_0 & = & E_2-E_0 + \frac{2\left(C_1 \cos\theta_+ + C_2 \sin\theta_-\right)^2}{E_2-E_0-\hbar\omega} \label{eq:E2minE0} \\
E'_3-E'_1 & = & E_3-E_1 + \frac{2\left(C_1 \cos\theta_+ - C_2 \sin\theta_-\right)^2}{E_2-E_0-\hbar\omega} \label{eq:E3minE1}
\end{eqnarray}
\end{subequations}
and
\begin{subequations}
\begin{equation}
\bra{1,N}' \hat{H}_0 \ket{0,N+1}' =
-C_1 \sin\theta_+ + C_2 \cos\theta_- + \left(C_1 \sin\theta_+ + C_2 \cos\theta_- \right)
\frac{
  \left(C_1 \cos\theta_+\right)^2 - \left(C_2 \sin\theta_-\right)^2
}{(E_2-E_0-\hbar\omega)^2}
\label{eq:1H0}
\end{equation}
\begin{equation}
\bra{3,N-1}' \hat{H}_0 \ket{2,N}' =
C_1 \sin\theta_+ + C_2 \cos\theta_- + \left( -C_1 \sin\theta_+ + C_2 \cos\theta_- \right)
\frac{
  \left(C_1 \cos\theta_+\right)^2 - \left(C_2 \sin\theta_-\right)^2
  }{(E_2-E_0-\hbar\omega)^2}.
\end{equation}
\end{subequations}

The above equations can now be used to determine the modified
conditions for performing the desired operation. One requirement for
suppressing the $\ket{0,N+1} \leftrightarrow \ket{1,N}$ oscillations is that
the parameters $C_1$, $C_2$ and $\omega$ are chosen such that
the matrix element $\bra{1,N}' \hat{H}_0 \ket{0,N+1}'$ [equation~(\ref{eq:1H0})] vanishes.
And in order to obtain full oscillations between the states
$\ket{2,N}$ and $\ket{3,N-1}$, the driving frequency $\omega$ must
match the energy difference $E'_3-E'_2$ [equation~(\ref{eq:E3minE2})].
These equations can be solved iteratively starting from the
weak-driving relations and calculating corrections. Note however that
at some level of accuracy one would need to take into account the coupling
between the different blocks in the dressed-state ladder.

It is however clear from the above results that strong driving does
not simply result in harmless modifications to the required
driving frequency and the ratio between the two driving
amplitudes. We identify three processes influencing the performance of the gate.
The first effect is related to the energy shifts in the
transitions with splittings \( E_2'-E_0' \) and \( E_3'-E_1' \) [equations~(\ref{eq:E2minE0}) and~(\ref{eq:E3minE1})]. They contain a common shift of \( 2[(C_1\cos\theta_+)^2+(C_2\sin\theta_-)^2]/(E_2-E_0-\hbar\omega) \). This shift represents the ac-Stark shift of the control qubit. For small \( \theta_+ \) and \( \theta_- \) we get \( 2C_1^2/(E_2-E_0-\hbar\omega) \), which is the analogue of the ac-Stark shift calculated in subsection~\ref{sec:acStark} (an estimate for this shift using typical experimental parameters is provided there).

Both the control qubit and target qubit transitions also get shifted with an amount that
is opposite for the two associated transitions: \( \pm 4 C_1 C_2 \cos\theta_+\sin\theta_-/(E_2-E_0-\hbar\omega) \).
What this result means is that during the operation of the CNOT
gate, an effective $\hat{\sigma}_z^{(1)} \hat{\sigma}_z^{(2)}$
term appears, which induces a (partial) CPHASE gate, and prevents us from obtaining
an ideal CNOT gate. This error is the equivalent of the CPHASE-type error discussed in
subsection~\ref{sec:acStark}.
If we take $J\ll\Delta_1-\Delta_2$ and the case of strong driving
(i.e.~$C_1\sim\Delta_1-\Delta_2$) we find that the undesirable
$\hat{\sigma}_z^{(1)} \hat{\sigma}_z^{(2)}$ term has a
coefficient that is of the order of $J^2/(\Delta_1-\Delta_2)$,
which causes small errors on the timescale of a CNOT gate, which
is given by $J^{-1}$. As discussed earlier, CPHASE-type errors can also be corrected
with single-qubit gates.

Lastly, when the matrix elements $H_{02}$, $H_{20}$, $H_{13}$ and $H_{31}$
can not be neglected, it is not possible for any pair of energy levels to be degenerate,
which is a general rule of quantum mechanics.
As a result no transition can be completely darkened in
this case, regardless of the choice of the parameters $C_1$, $C_2$
and $\omega$. To first approximation this source of errors can be regarded as performing
an additional single-qubit rotation, which is relatively easy to correct
for in experiments.

The results provided in this appendix can be used to perform a detailed numerical study
for optimizing the different parameters in order to maximize the fidelity of the CNOT
gate in the case of strong driving. However, we shall not do that here.

\end{document}